\PassOptionsToPackage{pdfpagelabels=false}{hyperref}
\documentclass[useAMS,usenatbib]{mnras}
\pdfoutput=1

\usepackage{newtxtext,newtxmath}
\usepackage{newtxtext,newtxmath}

\usepackage[T1]{fontenc}
\usepackage{ae,aecompl}

\usepackage{amsmath}	
\usepackage{siunitx}
\usepackage[utf8]{inputenc}
\usepackage{xcolor}
\usepackage{times}
\usepackage{color}
\usepackage[pdftex]{graphicx}
\usepackage{natbib}
\usepackage{comment}
\usepackage[normalem]{ulem}
\usepackage{xspace}

\def\hi{\ifmmode {\mbox H{\scshape i}}\else H{\scshape i}\fi\xspace}
\def\hii{\ifmmode {\mbox H{\scshape ii}}\else H{\scshape ii}\fi\xspace}
\def\h2{\ifmmode {\mbox H$_2$}\else H$_2$\fi\xspace}
\def\micron{\ifmmode {\mbox $\mu$m}\else $\mu$m\fi\xspace}

\title[The dust-continuum size of TNG50 galaxies at $z=1-5$]{The dust-continuum size of TNG50 galaxies at $z=1-5$: \\a comparison with the distribution of stellar light, stars, dust and \h2}

\author[G. Popping et al.]{Gerg\"o Popping$^{1}$\thanks{E-mail: gpopping@eso.org},
Annalisa Pillepich$^{2}$, Gabriela Calistro Rivera,$^{1}$ Sebastian Schulz$^{3}$, \newauthor Lars Hernquist$^{4}$, Melanie Kaasinen$^{2,5}$, Federico Marinacci$^{6}$, Dylan Nelson$^{5,7}$\newauthor  and Mark Vogelsberger$^{8}$
\\
\\
$^{1}$European Southern Observatory, Karl-Schwarzschild-Str. 2, D-85748, Garching, Germany\\
$^{2}$Max-Planck-Institut für Astronomie, Königstuhl 17, D-69117 Heidelberg, Germany\\
$^{3}$Institute for Computational Science, University of Zurich, Winterthurerstrasse 190, 8057 Zurich, Switzerland\\
$^{4}$Harvard-Smithsonian Center for Astrophysics, 60 Garden Street, Cambridge, MA, 02138, USA\\
$^{5}$Universit\"at Heidelberg, Zentrum f\"ur Astronomie, Institut f\"ur Theoretische Astrophysik, Albert-Ueberle-Straße 2, D-69120 Heidelberg, Germany\\
$^{6}$Department of Physics and Astronomy "Augusto Righi", University of Bologna, via Gobetti 93/2, 40129 Bologna, Italy\\
$^{7}$Max-Planck-Institut f\"u Astrophysik, Karl-Schwarzschild-Str. 1, D-85748 Garching, Germany\\
$^{8}$Kavli Institute for Astrophysics and Space Research, Department of Physics, MIT, Cambridge, MA, 02139, USA\\
}

\date{Accepted XXX. Received YYY; in original form ZZZ}

\pubyear{}

\begin{document}
\maketitle

\begin{abstract}
We present predictions for the extent of the dust-continuum emission of thousands of main-sequence galaxies drawn from the TNG50 simulation between $z=1-5$. To this aim, we couple the radiative transfer code \texttt{SKIRT} to the output of the TNG50 simulation and measure the dust-continuum half-light radius of the modeled galaxies, assuming a Milky Way dust type and a metallicity dependent dust-to-metal ratio. The dust-continuum half-light radius at observed-frame 850 \micron is up to $\sim$75 per cent larger than the stellar half-mass radius, but significantly more compact than the observed-frame 1.6 \micron (roughly corresponding to H-band) half-light radius, particularly towards high redshifts: the compactness compared to the 1.6 \micron emission increases with redshift. This is driven by obscuration of stellar light from the galaxy centres, which increases the apparent extent of 1.6 \micron disk sizes relative to that at 850 \micron. The difference in relative extents increases with redshift because the observed-frame 1.6 \micron emission stems from ever shorter wavelength stellar emission. These results suggest that the compact dust-continuum emission observed in $z>1$ galaxies is not (necessarily) evidence of the buildup of a dense central stellar component. We also find that the dust-continuum half-light radius very closely follows the radius containing half the star formation in galaxies, indicating that single band dust-continuum emission is a good tracer of the location of (obscured) star formation. The dust-continuum emission is more compact than the \h2 mass (for galaxies at $z\geq 2$) and the underlying dust mass. The dust emission strongly correlates with locations with the highest dust temperatures, which do not need to be the locations where most \h2 and/or dust is located. The presented results are a common feature of main-sequence galaxies.
\end{abstract}




\section{Introduction} \label{Introduction}
In the last decade, the Atacama Large (sub-)Millimeter Array (ALMA) has successfully observed the integrated dust-continuum emission of a few hundred of galaxies at $z>1$ \citep[and see also \citealt{Hodge2020} for a recent review]{Scoville2013, Scoville2014,Scoville2016, Scoville2017,Schinnerer2016,  Liu2019,Aravena2020}. One of the current challenges is to resolve the dust-continuum emission in high-redshift objects.  The dust-continuum emission is light from young stars that has been absorbed by dust and re-emitted at infrared (IR) and (sub-)millimeter wavelengths and is often used as a tracer of dust-obscured star formation. Resolved dust-continuum studies will thus provide important information about the location of dust-obscured star formation within galaxies and their link to the stellar buildup of galaxies over cosmic time. In particular, the distribution of dust-continuum emission from main-sequence galaxies (galaxies that lie on the observed correlation between the stellar mass and star formation rate (SFR) that the majority of star forming galaxies follow, \citealt{Noeske2007, Daddi2008,Whitaker2014}) is still largely unclear.

In recent years, there have been a number of observational studies focusing on the dust-continuum morphology of $z>1$ galaxies. These studies typically focused on the Far Infrared (FIR) brightest objects (sub-millimeter galaxies, SMGs, \citealt{Hodge2015,Simpson2015,Chen2017, CalistroRivera2018, Gullberg2019, Hodge2019, Lang2019}). \citet{Fujimoto2017} measured the extent of the dust-continuum emission in a sample of galaxies drawn from the ALMA archive between $z=0$ and $z=6$, typically focusing on FIR-bright objects. \citet{EricaNelson2018} explored the dust-continuum morphology of a $z=1.2$ main-sequence galaxy.  \citet{Rujopakarn2019} and \citet{Kaasinen2020} resolved the dust-continuum emission of three and two main-sequence galaxies at $z\sim 2.5$ and $z\sim 1.5$, respectively. \citet{Barro2016} used ALMA to resolve the dust-continuum emission of six compact star forming galaxies at $z\sim2.5$, whereas \citet{Tadaki2017} studied the dust-continuum distribution in two massive H$_\alpha$-selected galaxies at $z\sim 2.2-2.5$.  Recently, \citet{Tadaki2020} explore the 850 \micron size of 85 massive ($M_* > 10^{11}\,\rm{M}_\odot$) galaxies at $1.9 < z < 2.5$. What all of these observations have in common is that the dust-continuum emission of these galaxies is more compact than their optical or Near Infrared (NIR) emission. This has been seen as evidence of the dust-obscured buildup of a compact central dense stellar component \citep[e.g.,][]{Barro2016, Tadaki2020}. This conclusion is further supported by the finding that the dust-continuum emission is also more compact than the stellar mass distribution \citep{Lang2019,Tadaki2020}. Although the aforementioned studies of the resolved dust-continuum emission have been crucial to build our current understanding of obscured star formation in galaxies and the buildup of stellar disks, the number of targets is still mostly limited to $z=1-2$ main-sequence galaxies, or to very massive galaxies, not representative of the full galaxy population on the main-sequence. 

Besides tracing the dust-obscured star formation, the dust-continuum emission of galaxies has also frequently been used as a tracer of the molecular hydrogen (\h2) content of galaxies \citep[e.g.,][]{Scoville2013, Scoville2014,Scoville2016, Scoville2017, Eales2012, Bourne2013, Groves2015, Hughes2017, Schinnerer2016,Tacconi2018,Magnelli2020}. More detailed studies of individual objects have demonstrated that the \h2 mass of galaxies calculated from their integrated dust-continuum emission is similar to \h2 masses obtained by using the more classical tracer: carbon monooxide (CO) emission lines \citep{Hughes2017,Popping2017, Bertemes2018,Kaasinen2019, Aravena2020}. Theoretical efforts have reached similar conclusions \citep{Liang2018,Privon2018}. Extended configuration observations with ALMA have allowed a spatial comparison between various tracers of \h2 in galaxies. Observations of the brightest most actively star forming objects show that the dust-continuum emission is typically more compact than the CO emission \citep{Hodge2015, Chen2017,  CalistroRivera2018}. Using resolved observations of two main-sequence galaxies at $z\sim1.5$, \citet{Kaasinen2020} found that one of the objects has more extended CO emission than dust-continuum emission, whereas in the other the dust-continuum is more extended than the CO. 

It is expected that in the next years the extended baseline capabilities of ALMA will be used to increase the sample size and redshift coverage of main-sequence galaxies with resolved measurements of their dust-continuum emission (as well as CO and other lines). It is thus timely to develop the theoretical framework to yield predictions and put observations in a theoretical context, providing a physical explanation for the observations. In recent years a significant effort went into combining hydrodynamic galaxy-formation simulations with radiative transfer codes such as \texttt{SUNRISE} \citep{Jonsson2006}, \texttt{SKIRT} \citep{Baes2011,Baes2015,Camps2015} and \texttt{Powderday} \citep{Narayanan2020}. These tools have allowed theorists to provide direct predictions of the sub-mm dust-continuum emission of model galaxies, to be compared to observations from sub-mm observatories such as ALMA \citep[e.g.,][]{Camps2016,Behrens2018,Cochrane2019,Liang2020,Lovell2020}. The coupling of galaxy formation simulations with such radiative transfer codes is thus an ideal approach towards providing theoretical insights into the dust-continuum morphology of galaxies.

\citet{Cochrane2019} applied radiative transfer calculations with \texttt{SKIRT} on 4 galaxies modeled with the \texttt{FIRE-2} simulation \citep{Hopkins2018}, following these galaxies from $z=5$ to $z=1$, to study their resolved dust-continuum emission. The authors find that the simulated galaxies have dust-continuum emission that is generally more compact than the cold gas and the dust mass, but more extended than the stellar component. This study marked the first attempt to modeling the dust-continuum morphology of galaxies, but because of its limited sample size, it is hard to draw conclusions about main-sequence galaxies covering a range in galaxy properties (stellar mass, SFR, redshift). One of the reasons is that studies of the resolved simulated properties of galaxies require a high-mass resolution (with baryonic mass resolution elements of a few times $10^4\,\rm{M}_\odot$ or below) to sufficiently resolve galaxy disks. These simulations are computationally expensive to run for large cosmological volumes probing a wide galaxy parameter space.

The new TNG50 simulation (\citealt{Nelson2019}, \citealt{Pillepich2019}) is the highest resolution variant of the IllustrisTNG simulation suite (\citealt{Marinacci2018}, \citealt{Naiman2018}, \citealt{Nelson2018}, \citealt{Pillepich2018b}, \citealt{Springel2018}). With a mass resolution of $M_{\rm baryons} \sim 8.5 \times 10^{4}\,\rm{M}_\odot$ in a cosmological volume of $\sim$ 50 cMpc on a side, this simulation is ideal to study a representative sample of galaxies over cosmic time at sufficiently good mass resolution to resolve the inner structure of galaxy disks. 

In this work we couple the TNG50 simulation with the radiative transfer code \texttt{SKIRT} to make predictions for the dust-continuum emission of main-sequence galaxies at $z=1-5$. This redshift range was chosen to roughly match the observational data \citep[e.g.][]{Hodge2013,Karim2013,Fujimoto2017,Stach2019,Kaasinen2020} and to guarantee a decent number of main-sequence galaxies that are sufficiently resolved in the simulation. We specifically aim to address how the dust-continuum size of main-sequence galaxies evolves over cosmic time, if the dust-continuum emission is more compact than the stellar component and the optical/NIR emission of galaxies, and how well the dust-continuum emission of galaxies correlates with the distribution of \h2, dust mass and star formation in galaxies. We furthermore explore how the size of galaxies changes as a function of sub-mm wavelengths. 

The structure of the paper is as follows. In Section \ref{Methods} we present the methodology by introducing the TNG50 simulation and the \texttt{SKIRT} radiative transfer code, as well as the galaxy selection and our approach to measuring the sizes of the modeled galaxies. In Section \ref{sec:Results} we present the predictions by the model and we discuss these in Section \ref{sec:Discussion}. A summary and conclusion of the results presented in this work is given in Section \ref{sec:Conclusions}. Throughout this work we mostly focus on the 850 \micron size of galaxies (although see Section \ref{ref:wavelength}). The observed-frame emission at 850 \micron is covered by ALMA band 7 and is frequently used to study the dust-continuum emission of galaxies \citep[e.g.,][]{Hodge2013,Barro2016, Chen2017,Stach2019,Tadaki2020}.

\section{Description of the model} \label{Methods}
\subsection{The TNG50 Simulation} \label{TNG50}
In this paper we make use of the TNG50 simulation \citep{Nelson2019, Pillepich2019}. This simulation is part of the IllustrisTNG Project \citep[TNG hereafter:][]{Nelson2018, Marinacci2018, Springel2018, Naiman2018, Pillepich2018b}, a suite of magneto--hydrodynamical cosmological simulations for the formation of galaxies employing the moving mesh \texttt{AREPO} \citep{Springel2010} code.  The IllustrisTNG simulation is a revised version of the Illustris galaxy formation model \citep{Vogelsberger2013, Torrey2014}. The IllustrisTNG simulation evolves cold dark matter (DM) and gas from early times to $z=0$ by solving for the coupled equations of gravity and magneto-hydrodynamics (MHD) in an expanding Universe. The simulation includes prescriptions that describe the formation of stars, stellar evolution including the return of mass and metals from stars to the interstellar medium (ISM), gas cooling and heating, feedback from stars and feedback from supermassive black holes \citep[details are given in][]{Weinberger2017, Pillepich2018a}. Dark matter haloes and their substructures are identified using the \texttt{SUBFIND} algorithm outlined in \citet{Springel2010}.

The TNG50 simulation is the highest--resolution implementation of the TNG project. This simulation follows the evolution of $2 \times 2160^{3}$ intitial resolution elements inside a cube measuring 35 cMpc/h on each side, corresponding to a volume of 51.7$^{3}$ cMpc$^{3}$. This translates to a target mass resolution of  $0.85 \times 10{^5} M_\odot$ for the baryonic resolution elements (gas cells and stellar particles), $4.5 \times 10{^5} M_\odot$ for the dark matter resolution elements, a collisionless softening of about 0.3 kpc at $z=$ 0, and a minimum gas softening of 74 comoving parsecs.  The combination of a cosmological volume with a mass resolution of $0.85 \times 10{^5} M_\odot$ for the baryonic elements makes the TNG50 simulation perfectly suited to study the sizes of galaxies \citep[e.g.,][]{Pillepich2019}. The TNG50 simulation adopts a cosmology with $\Omega_{m} = 0.31,\,\Omega_{\Lambda} = 0.69,\,\Omega_{b} = 0.0486,\,h = 0.677,\,\sigma_8 = 0.8159,\,n_s = 0.97$, all consistent with the \citet{Planck2016} results. 
  
\subsection{Radiative transfer calculations with SKIRT}  
\label{sec:SKIRT}
To model the sub-mm emission of the galaxies from the TNG50 simulation we make use of the radiative transfer code \texttt{SKIRT} \citep{Baes2011, Camps2013,Saftly2014,Camps2015}. This code traces the scattering and absorption of photon packages by dust until they reach a detector. The code not only calculates an integrated flux for each simulated galaxy (as was for instance explored by, \citealt{Schulz2020} and \citealt{Vogelsberger2019} for IllustrisTNG simulations), but also generates a resolved image of the simulated galaxy at a predefined wavelength and resolution \citep[e.g.,][]{Rodriguez2019, Cochrane2019}. Here we largely adopt the same methodology for the radiation transfer calculations as \citet{Schulz2020}, which provides additional details. Below we summarize the methodology and highlight the parameters that were chosen differently from \citet{Schulz2020}.

To perform the radiative transfer calculations we extract the gas cells and stellar particles of the subhalo of interest. To achieve a face-on projection of every subhalo, we then rotate the coordinate systems of the gas and stellar cells/particles to align with the angular momentum vector of the gas in the subhalo (on average the face-on projection results in galaxy sizes 10 per cent more extended than a random projection). For stars older than 10 Myr, \texttt{SKIRT} calculates the spectral energy distribution (SED) of stellar particles based on the \citet{Bruzual2003} stellar population models according to their ages and metallicities.  The SED of stars younger than 10 Myr are modeled using the \texttt{MAPPINGS-III} photo-ionization code \citep[implemented internally within \texttt{SKIRT}]{Groves2008}, to account for the fact that young stars are located within their respective birth clouds. The \texttt{MAPPINGS-III} photo-ionization code includes emission from HII-regions, their surrounding photodissociation-regions (PDRS) and absorption by gas and dust in the birth clouds around young stars. We adopt the same assumptions as \citet{Schulz2020} when calculating the SED of young stars using the MAPPINGS-III code. Our results remain qualitatively the same when altering the parameters that describe the birth clouds that were given in \citet{Schulz2020}, or when not including birth clouds at all (the latter results in an increase of the dust-continuum radii increase of $\sim 10$\%). Throughout we consider all the gas cells and stellar particles located within 7.5 times the stellar half-mass radius of every subhalo.

The IllustrisTNG simulation suite does not directly follow the dust-abundance of gas cells. Instead, we adopt a dust-to-metal mass ratio (DTM) for gas cells that scales as a function of gas-phase metallicity. In particular, we adopt the relation between DTM and gas-phase metallicity derived for local galaxies by \citet[for a metallicity dependent CO-to-\h2 conversion factor, see also \citealt{DeVis2019}]{RemyRuyer2014}. \citet{RemyRuyer2014} found that DTM quickly increases as a function of metallicity up to a metallicity of 0.4 times solar. At higher metallicities the DTM takes a constant value of $0.32$. We assume that this relation is redshift independent, motivated by absorption studies of the dust-to-metal ratio through damped Ly-alpha and gamma-ray absorbers up to $z=5$ \citep{DeCia2016,Wiseman2017,Peroux2020}, as well as emission line studies of the dust-to-gas ratio of $z\sim 2$ galaxies \citep{Shapley2020}. The assumption of a redshift-independent relation between dust-to-metal ratio and gas-phase metallicity is also supported by some simulations \citep[cf.  \citealt{Hou2019}]{Popping2017dust, Li2019}. A dust abundance is only ascribed to gas cells with a temperature less than 75000 K or particles that are forming stars. We use a \citet{Weingartner2001} Milky Way Dust prescription to model a mixture of graphite, silicate and PAH grains. We do not account for dust self-absorption, but checked on a sub-sample of galaxies that the inclusion of dust self-absorption results in radii that are different by only $\sim 3$ per cent for main-sequence galaxies compared to our fiducial model setup. In Appendix \ref{sec:appendixDTM} we explore how the choice of the DTM relation influences our results.

As a part of the radiative transfer calculations, \texttt{SKIRT} also provides the dust temperature of every gas cell (see the examples in Figure \ref{fig:poststamps}). This dust temperature corresponds to the average temperature of grains of different species (composition and size) within a cell. This is different a priori from the dust temperature that is observationally derived through SED fitting (see \citealt{Liang2019} for a detailed discussion of different definitions of dust temperature). The CMB acts as an additional heating source of the dust and as a background against which the dust-continuum emission of the modeled galaxies is observed \citep{daCunha2013}. To account for this, we included the effects of the cosmic-microwave background (CMB) following \citet{Behrens2018} and \citet{Liang2019}. 

For the \texttt{SKIRT} output we define a wavelength grid of 100 uniformly spaced discrete wavelengths between the rest-frame UV and FIR wavelengths (in log space from 1000 Angstroms to 1 millimeter). The \texttt{SKIRT} output images at every wavelength are stored in fits images containing 450 pixels on a side. 

\subsection{TNG50 Galaxy Selection} 
\label{Galaxy Selection}
In this work, we focus on the dominant population of star forming galaxies, i.e. those that lie on the star formation main-sequence (SFMS) and above. We use the same classification of SFMS galaxies as \citet{Schulz2020}, based on the classification used in \citet[Section 4]{Pillepich2019}. \citet{Pillepich2019} calculated the median specific star formation rate (sSFR) inside twice the 3D stellar half mass radius of galaxies in 0.2 dex wide stellar mass bins running from  $10^8$ to $10^{10.2}\,\rm{M}_\odot$. All galaxies with a sSFR with a logarithmic distance (i.e., $\log{\rm sSFR} - \log{\rm{sSFR}_{\rm median}}$) less than -0.5 from the median sSFR were rejected. The process was then repeated until the median converged. A powerlaw was then fitted to the sSFR median to extrapolate the main-sequence to higher stellar masses. All galaxies with a logarithmic distance larger than -0.5 were then classified as being on the main-sequence or above. We do not make a distinction between central and satellite galaxies. For this work we focus on stellar masses and star formation rates within twice the stellar half-mass radius of every subhalo.

We do not include galaxies with a stellar mass less than $10^9\,\rm{M}_\odot$. This is done to ensure that the individual galaxies are resolved well enough to study their morphological properties. In this work we focus on galaxies with redshifts $z \leq 5$ to guarantee a decent number of galaxies that meet our initial selection criteria at high redshifts. An overview of the number of galaxies considered in this study is given in Table \ref{tab:selection}.

\subsection{Quantifying the size of galaxies}
Throughout this paper the size of a galaxy corresponds to the half-light or half-mass radius of the galaxy. In particular, when we discuss the dust-continuum size of a galaxy we refer to the radius that contains half the light emitted at the respective wavelength. The same is true when we discuss the 1.6 \micron size, which corresponds to the radius within which half the emission at an observed wavelength of 1.6 \micron is located. When we discuss the stellar, \h2 or dust size of galaxies we refer to the radius within which half the stellar, \h2 or dust mass is located, respectively. Lastly, when we discuss the SFR size of a galaxy, we refer to the radius within which half the instantaneous star formation of a galaxy takes place. Consistently with the case of dust-continuum, we measure the stellar, \h2, dust, and SFR size of the model galaxies by adopting a face-on projection and only by accounting for gas cells/stellar particles within 7.5 times the stellar half-mass radius. To calculate the \h2 mass of a gas-cell in the TNG50 simulation we made use of the methodology presented in \citet{Popping2019}, adopting the \citet[GK]{Gnedin2011}, \citet[K13]{Krumholz2013} and \citet[BR]{Blitz2006} \h2 recipes (see \citealt{Popping2019} for more details). The dust mass corresponds to the metal mass in the ISM multiplied by the dust-to-metal ratio set for every cell as a function of the gas-phase metallicity (see Section \ref{sec:SKIRT}).

We acknowledge that the adopted definition for galaxy size may be different from observationally defined sizes, such as the radius within which 80\% of the light is contained, or for example the scale radius obtained when fitting an exponential function to the light profile. Furthermore, we have not attempted to account for instrumental effects such as PSF/beam properties and limiting sensitivity. Because these effects are not included we refrain ourselves from a quantitative comparison with observations both in the figures and in writing.

\subsection{Examples}
\begin{figure*}
\begin{center}
\includegraphics[width = 1.0\hsize]{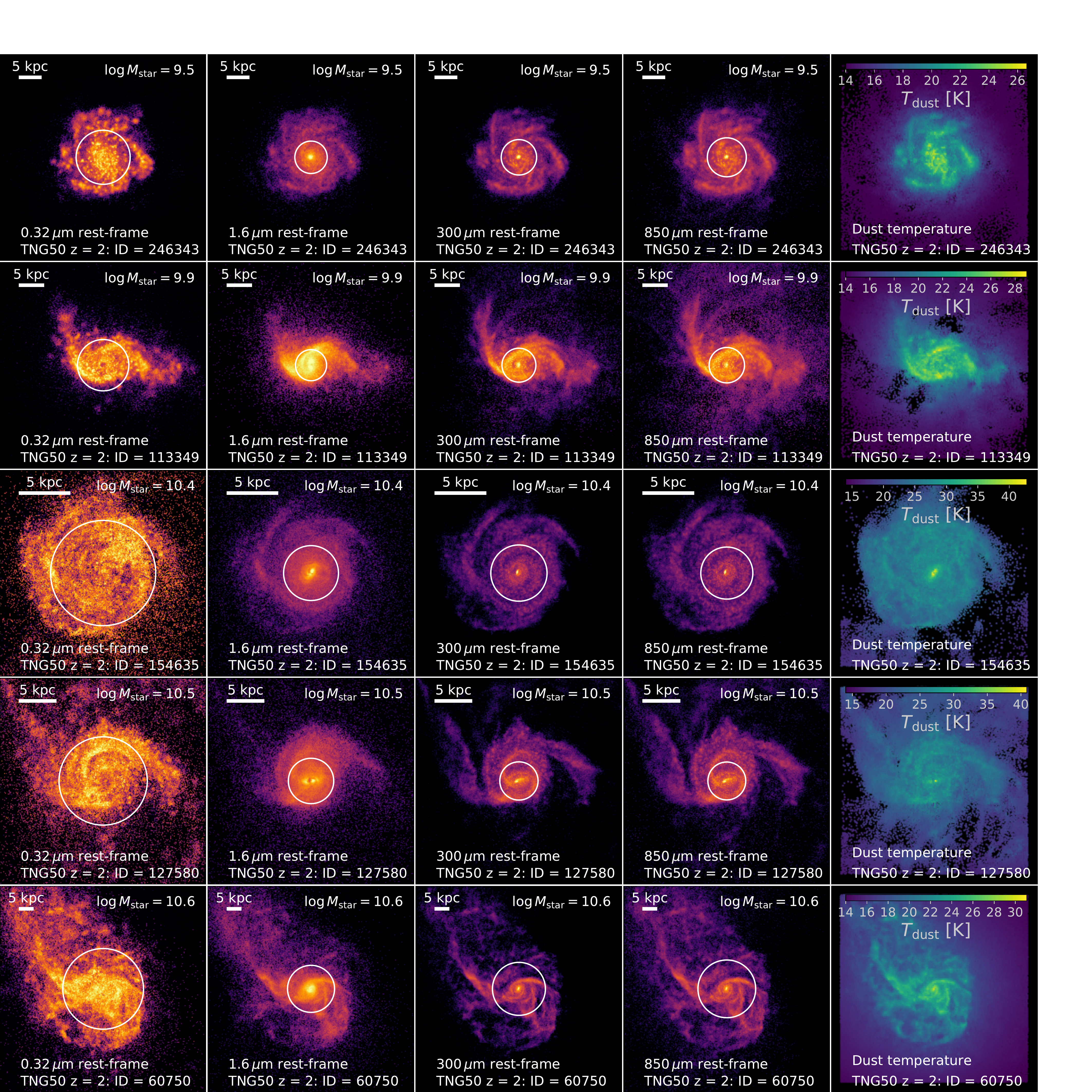}
\caption{Rest-frame 0.32 \micron~(left column), 1.6 \micron~(center left column), 300 \micron~(center column) and 850 \micron~(center right column) images of 5 example galaxies from TNG50 at $z=2$ with a range of stellar masses. White circles correspond to the respective half-light radii. The column at the right show the dust temperature distribution of these galaxies. The morphology and, of interest for this paper, the half-light radii of galaxies vary
across wavelengths. \label{fig:poststamps}}
\end{center}
\end{figure*}

By combining IllustrisTNG and \texttt{SKIRT} we can not only study the size of the gas components of modeled galaxies, but also the extent of the stellar and dust emission. In Figure \ref{fig:poststamps} we have included a number of example $z=2$ modeled galaxies at different wavelengths. We find that the appearance of the galaxies themselves changes as a function of wavelength. At a rest-frame wavelength of 0.32 \micron the emission is much more extended than at 1.6, 300 and 850 \micron, where we see more centrally concentrated bright emission. This is a reflection of stellar emission being absorbed and re-emitted. From these examples it also becomes clear that the morphology and, of interest for this paper, the sizes of galaxies vary across wavelengths. In the fifth column of Figure \ref{fig:poststamps} we show the dust temperature distribution of the example galaxies. We find a negative dust temperature gradient across the galaxy disk for all these examples. The negative gradient has implications for the dust-continuum size evolution across FIR wavelengths as we will discuss in the next section.

\begin{table}
  \centering
  	\begin{tabular}{cc} 
		\hline
		Redshift &  \# TNG50 galaxies:  \\
		$z$ & star forming and $M_* > 10^9\,\rm{M}_\odot$  \\
		\hline
        1 & 1710\\
        2 & 1149\\
        3 & 620\\
        4 & 206\\
        5 & 76\\
		\hline
	\end{tabular}
	\caption{An overview of the number of TNG50 galaxies used in this work.}
	\label{tab:selection}
\end{table}

\begin{figure*}
\includegraphics[width = 1.0\hsize]{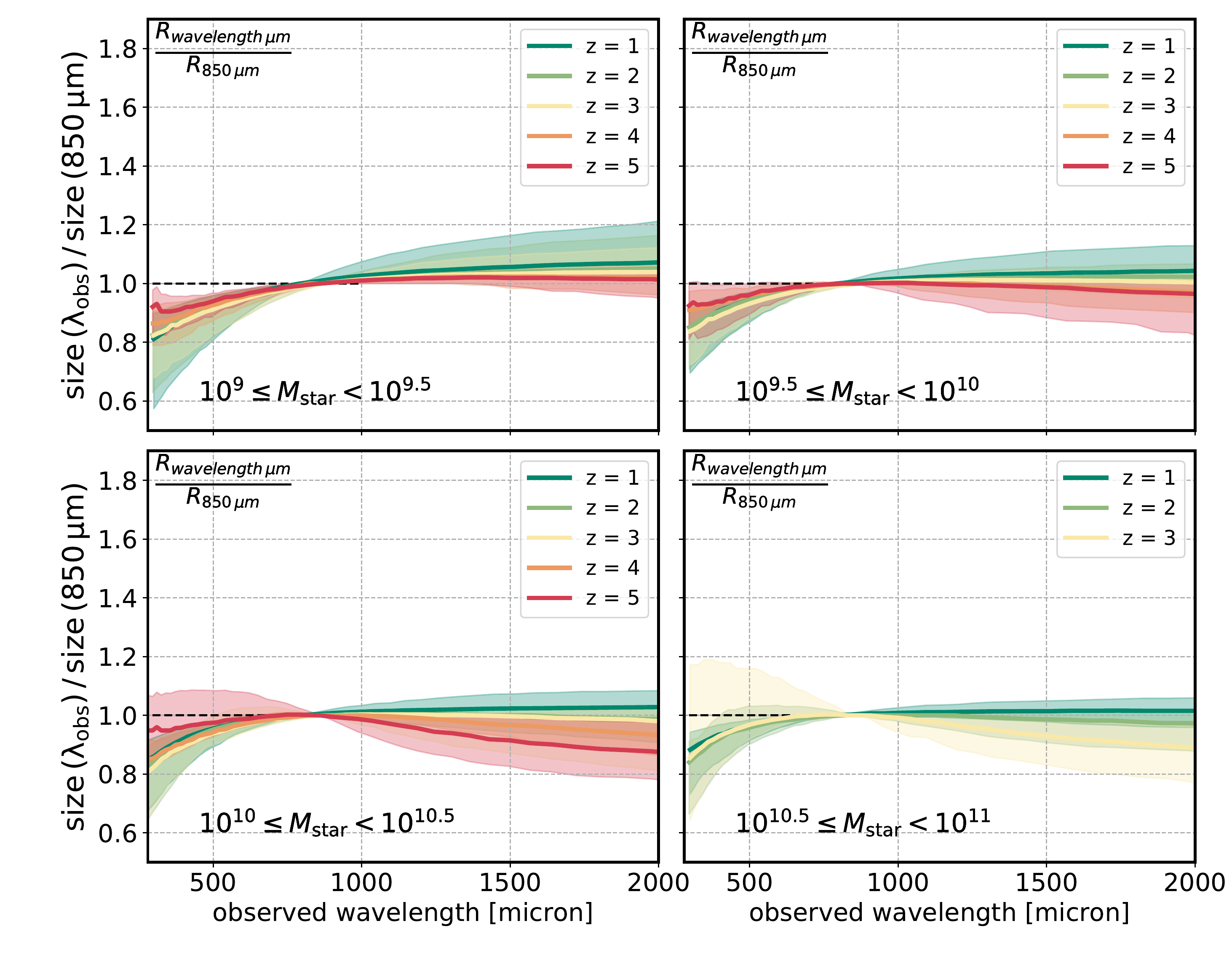}
\caption{The ratio between the dust-continuum half-light radius of the modeled galaxies at 850 \micron and other wavelengths as a function of observed-frame wavelength. The ratios are presented for a range of redshifts and the panels correspond to bins of stellar mass. The solid curves correspond to the median of the distribution, whereas the color-shaded regions mark the one-sigma scatter of the distribution. The black dashed horizontal line corresponds to a ratio of unity. The bottom-right panel does not include galaxies at $z=4$ and $z=5$, due to low numbers. The half-light radii of galaxies at wavelengths running from 700 \micron to 2 mm (ALMA bands 4 through 8) are similar at the $\lesssim 5-10$ per cent level.  \label{fig:wavelength}}
\end{figure*}

\section{Results}
\label{sec:Results}
In this section we present the dust-continuum size of galaxies and how this relates to other galaxy properties.

\subsection{The size of galaxies across the sub-mm wavelength regime}
\label{ref:wavelength}
Since the shape and peak location of the IR SED depend among others on the temperature distribution of the dust, one may expect differences in the dust-continuum size of galaxies as a function of observed wavelength. Furthermore, an observed wavelength of 850 \micron corresponds to a range in rest-frame wavelengths when looking at galaxies at increasing redshifts (e.g., at $z=4$ an observed wavelength of 850 \micron corresponds to a rest-frame wavelength of 170 \micron). To better quantify this we first present the dust-continuum size of galaxies for different observed-frame wavelengths.

Figure \ref{fig:wavelength} shows the ratio between the dust-continuum size of the modeled galaxies at an observed-frame wavelength of 850 \micron and the size at other wavelengths running from 300 \micron to 2 mm at different redshifts. These ratios are plotted at various redshifts and in bins of stellar mass. Due to the redshifting of the galaxy SED, a fixed observed wavelength probes shorter rest-frame wavelengths with increasing redshift. We find that size ratios are approximately unity compared to the 850 \micron sizes for dust-continuum sizes estimated in the wavelength range from 700 \micron to 2 mm. At shorter wavelengths the sizes are typically smaller than the 850 \micron size (with a gentle increase in the size ratio with increasing stellar mass).  For wavelengths shorter than 700 \micron, the size ratio compared to 850 \micron on average increases with increasing redshift, wheareas this size ratio on average decreases with increasing redshift for size estimates at wavelengths longer than 1 mm. The change in size with observed wavelength is similar to the findings by \citet{Cochrane2019}, who found that the dust-continuum size of a single modeled $z\sim3$ galaxy increases as a function of observed wavelength. We do not find any trend between the various bins of stellar mass.

To first order, the change as a function of wavelength is driven by the dependence of the sub-mm emission strength on the temperature of the dust. At rest-frame wavelengths longer than $\sim 850$ \micron, the Rayleigh-Jeans (RJ) tail, the dust-continuum emission is linearly proportional to the temperature of the dust. Below a rest-frame wavelength of $\sim 350$ \micron the dust becomes optically thick and the strength of the dust-continuum emission depends more strongly on the temperature of the dust. Thus, small deviations in the temperature across the disk will look more pronounced in the dust-continuum emission at a rest-frame wavelength of  200 \micron than at 850 \micron or 1 mm. A small negative dust temperature gradient across the disk therefore naturally translates into smaller sizes at wavelengths below the RJ tail, than at wavelengths above the RJ tail. Indeed, negative gradients in dust temperature across the galaxy disks are seen in Figure \ref{fig:poststamps}. In Figure \ref{fig:wavelength} we notice changes in disk size already at observed wavelengths of 600 \micron and below. This is a natural consequence of the change in the observed SED of galaxies as a function of redshift. At $z=1$, an observed-frame wavelength of 600 \micron already corresponds to a rest-frame wavelength of 300 \micron (i.e., below the RJ-tail), hence being more sensitive to changes in dust temperature. Similarly, an observed wavelength of 3 mm corresponds to a rest-frame wavelength of 600 \micron at $z=4$, still probing the Rayleigh-Jeans tail whereas the rest-frame 850 \micron emission probes emission shortward of the Rayleigh-Jeans regime. This naturally leads to a 3 mm size larger than the 850 \micron size.

Importantly, Figure \ref{fig:wavelength} also shows that the sizes of galaxies in the observed-wavelength regime running from 700 \micron to 2 mm are similar at the $\lesssim 5-10$ per cent level. This  is because the wavelengths are close enough to each other to be dominated by dust at predominantly the same temperatures. This is also encouraging, as these observed-wavelengths roughly correspond to ALMA bands 6 through 8, which are the bands most frequently used for (resolved) continuum studies of galaxies. Sizes estimated in these different bands can thus safely be compared to each other, after accounting for a systematic offset up to a few per cent.

\begin{figure}
\includegraphics[width = 1.0\hsize]{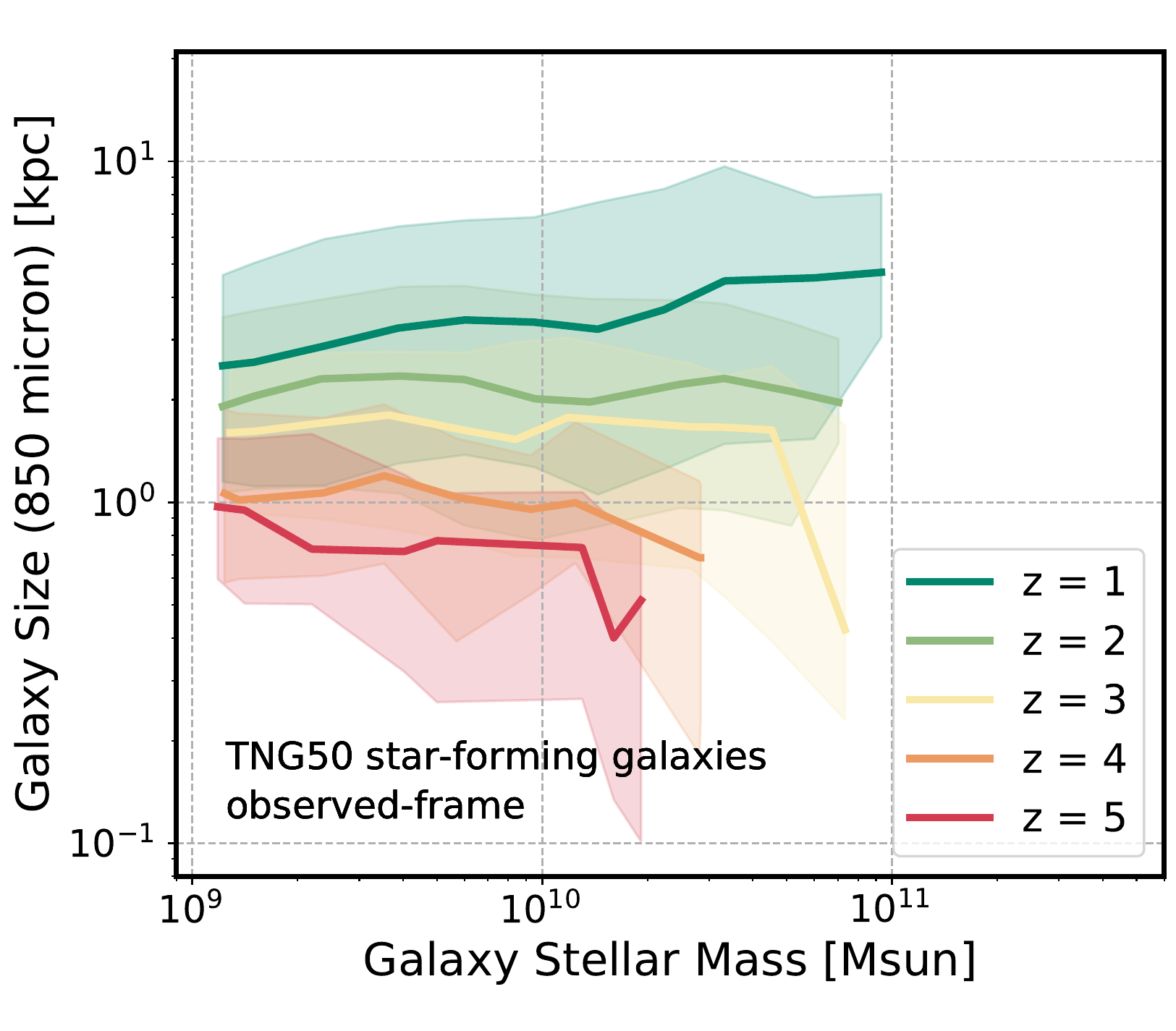}
\caption{The median dust-continuum size of galaxies at an observed wavelength of 850 \micron as a function of stellar mass at different redshifts, as predicted by TNG50+\texttt{SKIRT}, assuming a Milky Way dust type and a metallicity dependent dust-to-metal ratio. The model predicts an increase in dust-continuum size at fixed stellar mass with cosmic time. The solid curves correspond to the median of the distribution, whereas the color-shaded regions mark the one-sigma scatter of the distribution.  \label{fig:size_evolution}}
\end{figure}

\subsection{The stellar mass -- dust-continuum size relation of galaxies}
In Figure \ref{fig:size_evolution} we show the observed-frame 850 \micron size of galaxies as a function of their stellar mass for various redshifts. At $z=1$ the model predicts a very gentle increase in the 850 \micron size of galaxies as a function of stellar mass. At $z=2-4$ the 850 \micron size of galaxies is roughly constant as a function of stellar mass (except for the quick drop seen for the most massive galaxies at $z=3$, driven by small number statistics). At $z=5$ there appears to be a gentle decrease in the 850 \micron size of galaxies as a function of stellar mass. 

At fixed stellar mass the TNG50+\texttt{SKIRT} model predicts an increase in the observed 850 \micron size of galaxies as a function of cosmic time. The redshift evolution thus does not necessarily say something about the evolution of the actual dust distribution of galaxies, but rather also strongly depends on the temperature of the emitting dust. At $z=5$, the 850 \micron emission probes the peak of the FIR SED, whereas at $z=1$ the 850 \micron emission already probes the Rayleigh Jeans tail of the FIR SED with a different dependence on the dust temperature. 

\begin{figure*}
\includegraphics[width = 1.0\hsize]{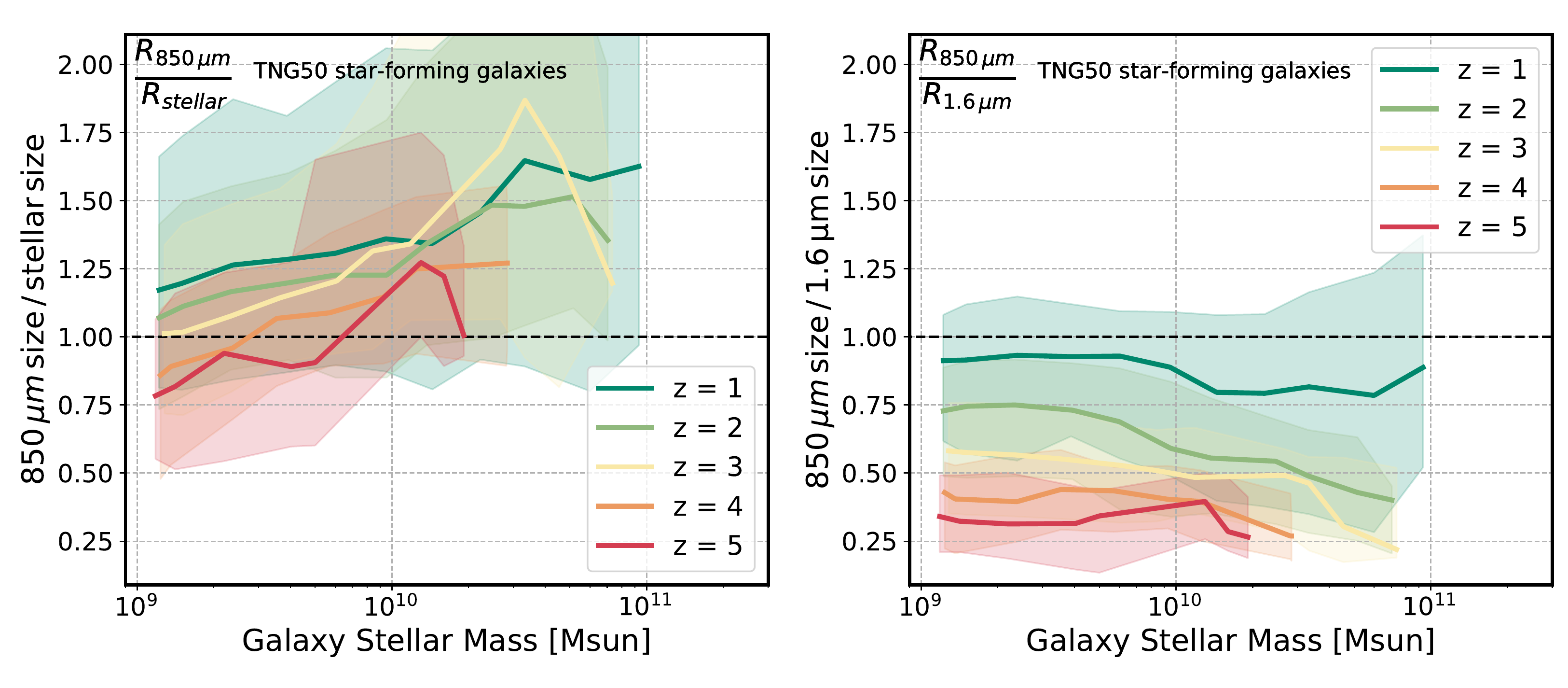}
\caption{Left panel: The ratio between dust-continuum size of galaxies at an observed wavelength of 850 \micron and the stellar half-mass radius as a function of stellar mass at different redshifts. Right panel: The ratio between dust-continuum size of galaxies at an observed wavelength of 850 \micron and the 1.6 \micron half-light radius as a function of stellar mass at different redshifts, the latter approximating the stellar light radial extent of galaxies in H-band.  The solid curves correspond to the median of the distribution, whereas the color-shaded regions mark the one-sigma scatter of the distribution. The black dashed horizontal line corresponds to a ratio of unity. The observed-frame 850 \micron emission is typically more extended than the stellar mass, but more compact than the observed-frame 1.6 \micron emission. \label{fig:StellarRatio}}
\end{figure*}

\subsection{The ratio between the dust-continuum size and stellar size of galaxies}
\label{sec:ratioStar}
In the left panel of Figure \ref{fig:StellarRatio} we show the ratio between the 850 \micron continuum half-light radius and the stellar half-mass radius of modeled galaxies as a function of their stellar mass. The model predicts an increase in this ratio as a function of stellar mass, from ratios around unity at stellar masses of $10^9\,\rm{M}_\odot$ up to $\sim 1.75$ at $10^{10.5}\,\rm{M}_\odot$. Namely, the dust-continuum half-light radius at observed-frame 850\micron is up to $\sim$75 per cent larger than the stellar half-mass radius and more so at lower rather than higher redshifts. Moreover, the galaxy-to-galaxy variations, indicated by the shaded areas at one-sigma scatter, are very large. At stellar masses smaller than $10^{10}\,\rm{M}_\odot$ we see a gentle increase in the ratio with redshift.

The prediction that the ratio between 850 \micron half-light and stellar half-mass radii is typically around or larger than one seems to be in conflict with current observations. However, the observations do not probe the stellar mass distribution directly, but are commonly measured using a NIR filter (for instance the K and H-band in \citealt{vanderWel2013}). 

To allow for a better comparison with the observations, we show the ratio between the 850 \micron and 1.6 micron observed-frame half-light radii of the modeled galaxies as a function of their stellar mass and redshift in the right panel of Figure \ref{fig:StellarRatio}. The model predicts a very gentle decrease in this ratio as a function of stellar mass at redshifts $z=2$ and $z=3$, whereas at the other redshifts the predicted ratio is roughly constant. Furthermore, the model predicts an increase in the ratio with cosmic time. Most importantly, the ratio between the 1.6 and 850 \micron half-light radii is around one at $z=1$ and decreases towards higher redshifts, down to ratios of 0.3 not 0.5, right? at $z=5$. This is in much better qualitative agreement with the observations than the predicted ratio between 850 \micron and stellar half-mass radius. In Section \ref{sec:dust_effect} we will discuss in more detail the reasons behind this different behavior.

\begin{figure*}
\includegraphics[width = 1.0\hsize]{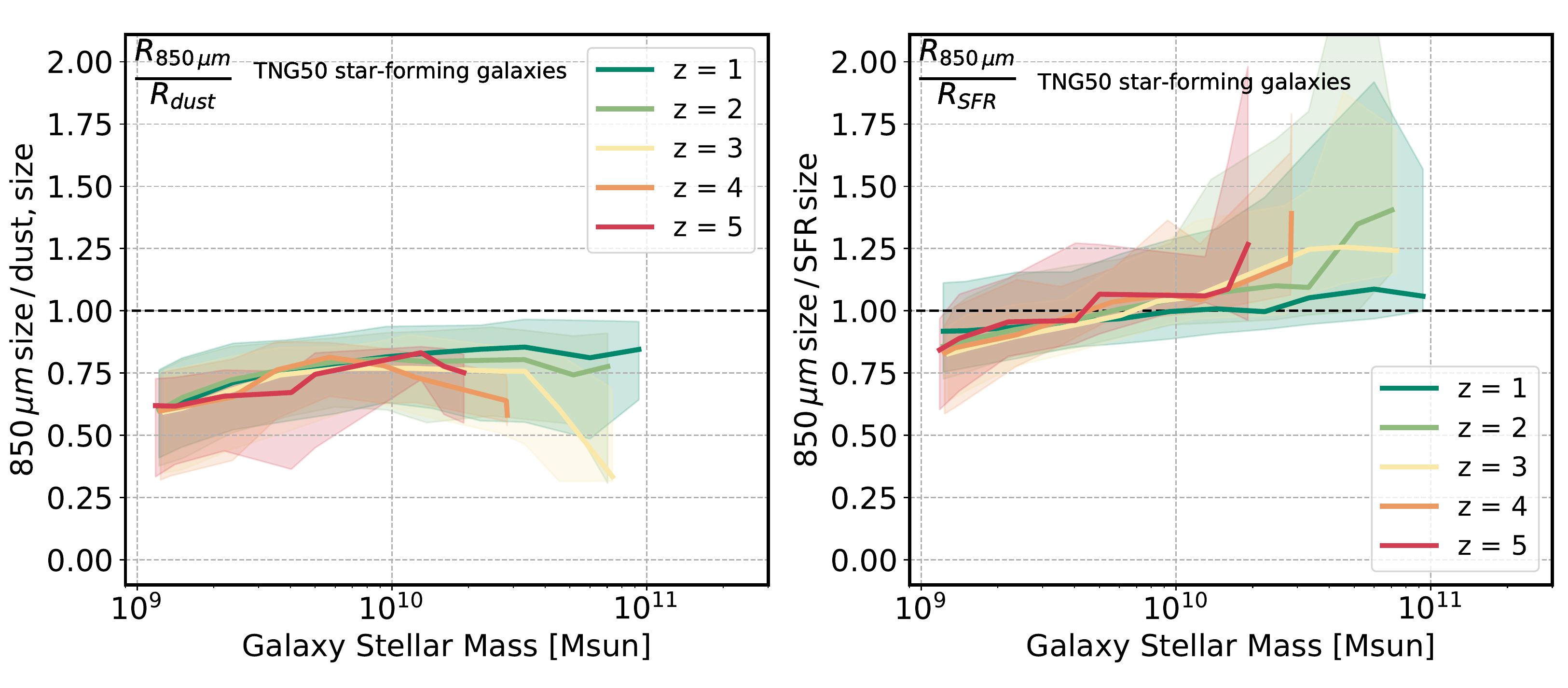}
\caption{Left panel: The ratio between dust-continuum half-light radius of galaxies at an observed wavelength of 850 \micron and the dust half-mass radius as a function of stellar mass at different redshifts. Right panel: The ratio between dust-continuum half-light radius of galaxies at an observed wavelength of 850 \micron and the radius containing half of the SFR as a function of stellar mass at different redshifts.  The solid curves correspond to the median of the distribution, whereas the color-shaded regions mark the one-sigma scatter of the distribution. The black dashed horizontal line corresponds to a ratio of unity. The dust-continuum half-light radius is more compact than the dust half-mass radius and is similar to the radius containing half the SFR.  \label{fig:SFRDustRatio}}
\end{figure*}

\subsection{The ratio between the dust-continuum size and dust size of galaxies}
In the left-hand panel of Figure \ref{fig:SFRDustRatio} we present the ratio between the observed-frame dust-continuum half-light radius and the dust half-mass radius of galaxies as a function of their stellar mass and redshift.  We find that the dust-continuum half-light radius is smaller than the dust half-mass radius, with a typical ratio of 0.75 independent of redshift and stellar mass. This suggests that the observed dust-continuum emission of galaxies at 850 \micron is not a perfect tracer of the underlying dust distribution, but rather underestimates (somewhat) the extent of the dust distribution.

\subsection{The ratio between the dust-continuum size and the SFR size of galaxies}
In the right-hand panel of Figure \ref{fig:SFRDustRatio} we show the predicted ratio between the observed-frame 850 \micron size of galaxies and the radius containing half the SFR as a function of stellar mass and redshift. We find that this ratio gently increases with stellar mass and has no significant redshift evolution. The median ratio runs from $\sim 0.85$ to $\sim 1.15$ in the stellar mass range from $10^9$ to $10^{10.8}\,\rm{M}_\odot$. This ratio is by far the closest to unity of all the other ratios explored in this work (also in the next subsection). This indicates that the resolved dust-continuum emission distribution of galaxies is very closely linked to the distribution of star formation within the galaxies. We will further discuss this in Section \ref{sec:discussion_dustcontinuumorigin}.

\begin{figure*}
\begin{center}
\includegraphics[width = 1.0\hsize]{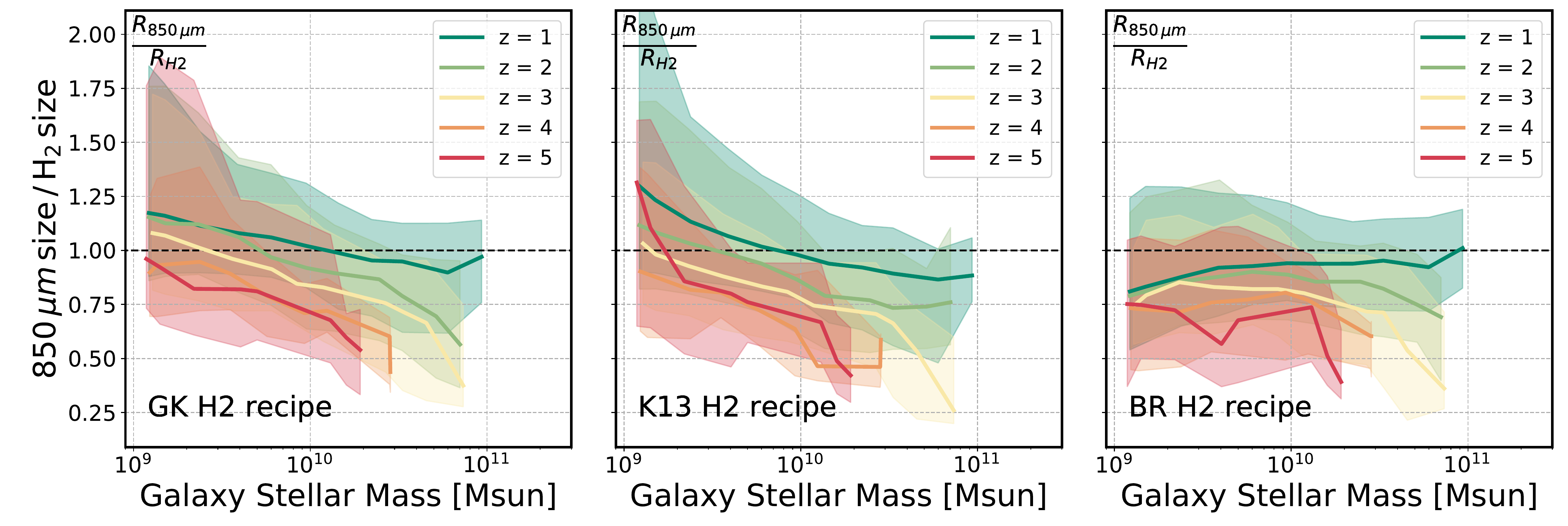}
\caption{The ratio between dust-continuum half-light radius of galaxies at an observed wavelength of 850 \micron and the molecular hydrogen half-mass radius as a function of stellar mass at different redshifts for the GK (\citealt{Gnedin2011}, left), K13 (\citealt{Krumholz2013}, middle) and the BR (\citealt{Blitz2006}, right) \h2 recipes.  The solid curves correspond to the median of the distributions, whereas the color-shaded regions mark the one-sigma scatter of the distributions. The black dashed horizontal line corresponds to a ratio of unity. Above stellar masses of $10^{10}\,\rm{M}_\odot$ the predicted ratios for $z\geq 2$ galaxies are typically below one (smaller than 0.75), independent of the adopted \h2 recipes. At lower stellar masses the ratios predicted when adopting the three \h2 recipes differ. \label{fig:H2Ratio}}
\end{center}
\end{figure*}

\subsection{The ratio between the dust-continuum half-light radius and \h2 half-mass radius}
In Figure \ref{fig:H2Ratio} we plot the ratio between the observed-frame 850 \micron size of galaxies and the \h2 half-mass size as a function of stellar mass and redshift when adopting the \citet[GK]{Gnedin2011}, \citet[K13]{Krumholz2013} and \citet[BR]{Blitz2006} \h2 recipes. We find that this ratio on average decreases with redshift, independent of stellar mass and adopted \h2 recipe. For all three recipes, at $z=1$ the ratio is typically close to one for galaxies with stellar masses larger than $10^{10}\,\rm{M}_\odot$. At $z=2$ the median ratio is already $\sim0.8$ at these stellar masses, whereas it decreases down to $\sim 0.5$ for galaxies with stellar masses of a few times $10^{10}\,\rm{M}_\odot$ at $z \geq 4$.

At stellar masses below $10^{10}\,\rm{M}_\odot$ the model predicted ratio between the observed-frame 850 \micron size of galaxies and the \h2 half-mass size is more dependent on the adopted \h2 recipe. When adopting the GK and K13 recipes, we find a gentle decrease in the ratio between 850 \micron size of galaxies and the \h2 half-mass size as a function of stellar mass. When adopting the BR  recipe the ratio is essentially constant with stellar mass, and we only see a redshift dependence. It is worthwhile to mention that both the GK and K13 recipes include a metallicity dependence, whereas the BR recipe does not. A metallicity gradient (with lower metallicities in the outskirts of TNG50 galaxies, see \citealt{Hemler2020}) thus naturally leads to changes in the \h2 half-mass size of the model galaxies.

Despite the difference between the \h2 recipes, these results suggest that the dust-continuum emission of galaxies is not necessarily a good tracer of the resolved \h2 distribution: especially at $10^{10}\,\rm{M}_\odot$ the 850 \micron size of the modeled galaxies is smaller than the \h2 half-mass size at $z\geq 2$. In a future work we aim to compare the dust-continuum emission to the CO emission of galaxies, allowing for a more direct comparison with observations.

\begin{figure*}
\includegraphics[width = 1.0\hsize]{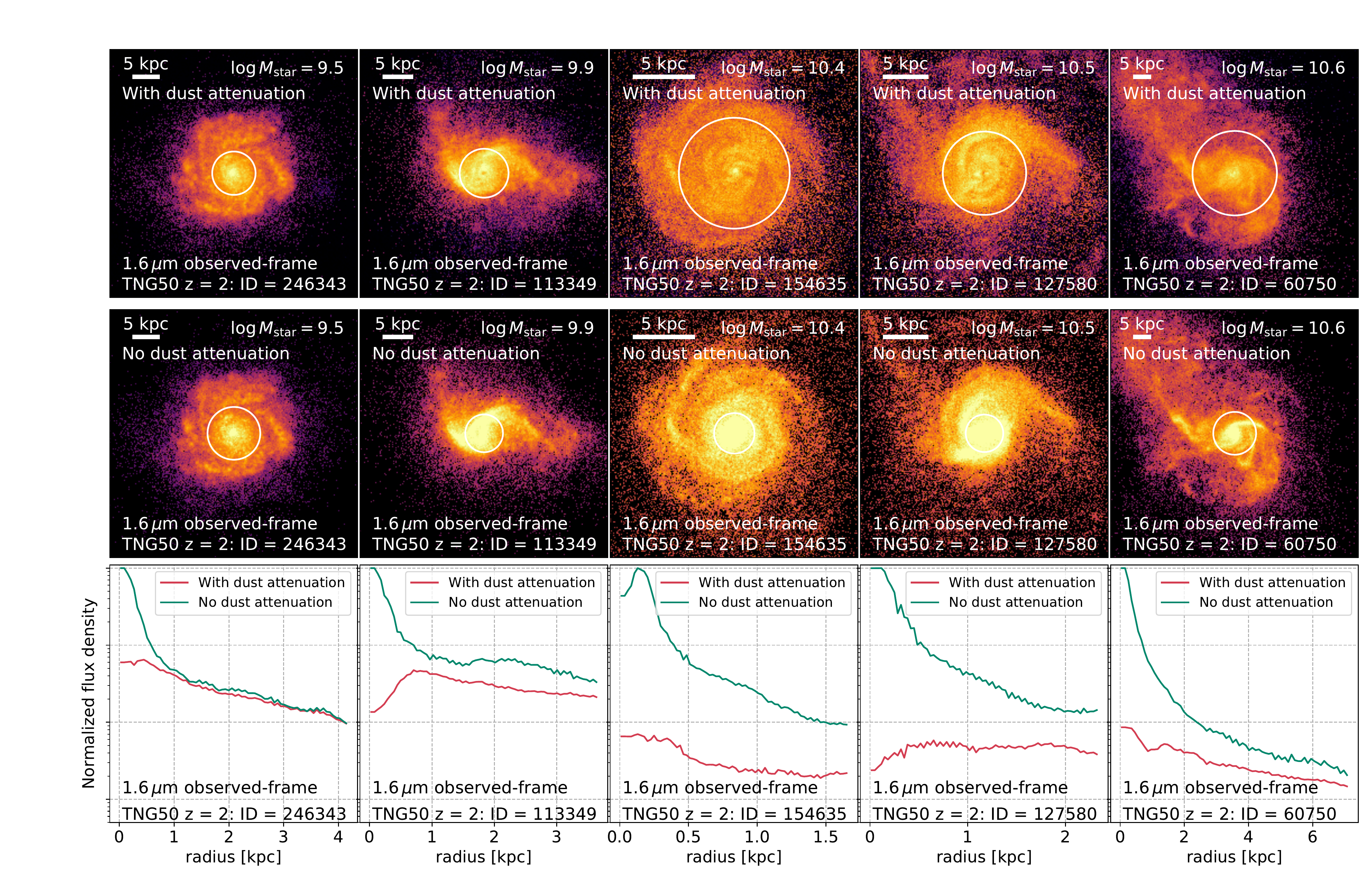}
\caption{Poststamps of the observed-frame 1.6 \micron emission of the example galaxies at $z=2$ displayed in Figure \ref{fig:poststamps}. The top row corresponds to a \texttt{SKIRT} run including the effects of dust attenuation, whereas the middle row corresponds to the scenario where no dust attenuation is applied. White circles correspond to the respective half-light radii. For individual galaxies, the top and middle row have the same color scaling. Especially for the more massive galaxies it becomes clear that dust obscures the central bright component of the stellar emission. The bottom row shows the 1.6 \micron light profile of the same galaxies when including (red) and not including (green) dust absorption. The radial profiles are normalized to the peak flux density of the profile without dust attenuation. The shape of the profiles is flatter when including dust attenuation, naturally extending the half-light radius.  \label{fig:HbandPoststamps}}
\end{figure*}

\begin{figure*}
\begin{center}
\includegraphics[width = 1.0\hsize]{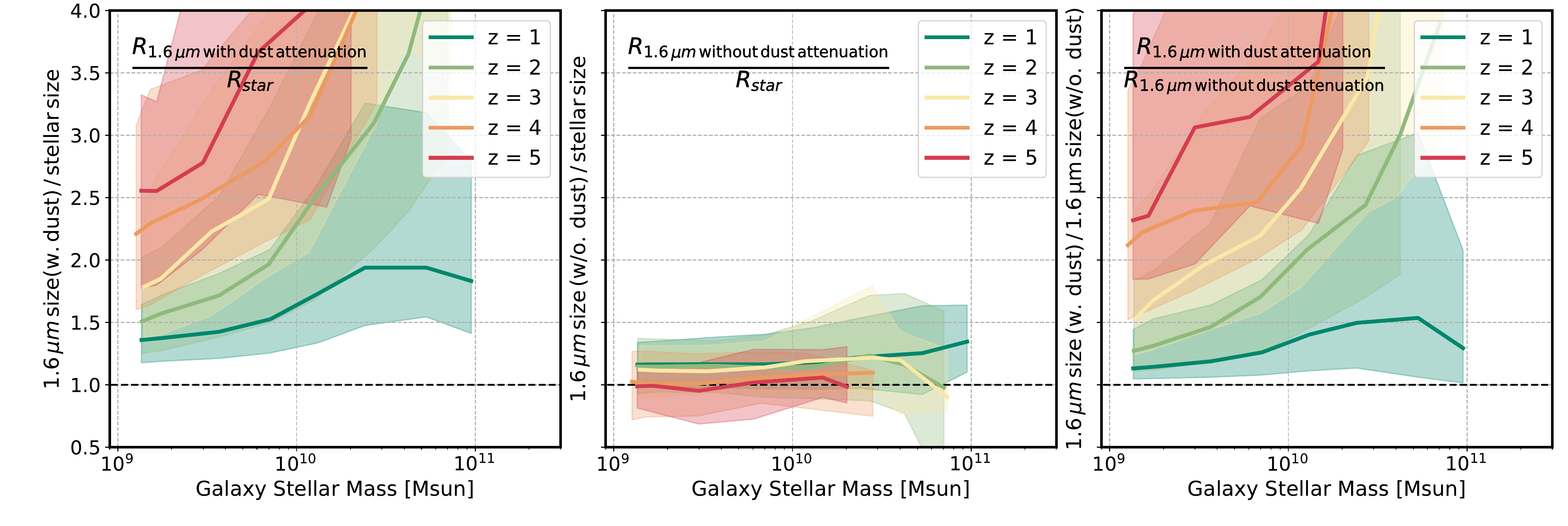}
\caption{Left: The ratio between observed-frame 1.6 \micron half-light radius of the modeled galaxies and their stellar half-mass radius as a function of stellar mass and redshift, including the effects of dust attenuation as in Figure \ref{fig:HbandPoststamps}. Center: The ratio between 1.6 \micron half-light radius of the modeled galaxies and their stellar half-mass radius as a function of stellar mass and redshift, not including the effects of dust attenuation. Right: The ratio between the 1.6 \micron half-light radius of galaxies with and without including dust attenuation plotted as a function of stellar mass and redshift.  The solid curves correspond to the median of the distribution, whereas the color-shaded regions mark the one-sigma scatter of the distribution. The black dashed horizontal line corresponds to a ratio of unity.\label{fig:HbandRatio}}
\end{center}
\end{figure*}

\section{Discussion}
\label{sec:Discussion}
\subsection{What drives the compact distribution of dust emission in galaxies?}
\label{sec:dust_effect}
In Section \ref{sec:ratioStar} we showed that the ratio between the dust-continuum half-light radius of galaxies and their stellar half-mass size is typically above unity. The ratio with the 1.6 \micron half-light radius of galaxies is typically below unity for galaxies at $z\geq2$ and decreases as a function of redshift, so that galaxies observed in dust-continuum appear progressively smaller than in H-band stellar light at progressively higher redshifts. The latter is in agreement with observations, that have also suggested that the dust-continuum sizes of galaxies are more compact than their observed NIR sizes \citep[e.g.,][]{Barro2016, Fujimoto2017, Tadaki2017,CalistroRivera2018,EricaNelson2018, Lang2019,Puglisi2019, Lang2019, Rujopakarn2019, Tadaki2020}. Here, we aim to give more insight into the origin of the compact dust emission compared to the 1.6 \micron emission.

An observed wavelength of 1.6 \micron corresponds to increasingly bluer rest-frame wavelengths as a function of redshift. As a result, the effect of dust-obscuration of the observed 1.6 \micron  emission becomes increasingly more important as the redshift increases. For example, at $z>3$ the observed 1.6 \micron emission already corresponds to the UV regime, in which the obscuration by dust plays an important role. The attenuation of stellar emission by dust may thus play an important role in determining the observed 1.6 \micron morphology of galaxies. To examine this in more detail we re-ran the radiative transfer models, but without including dust. As a consequence, the stellar emission is not absorbed by dust \emph{and} therefore no obscured emission is reemitted in the IR either. In Figure \ref{fig:HbandPoststamps} we show the same $z=2$ example galaxies as in Figure \ref{fig:poststamps}, but at an observed wavelength of 1.6 \micron (corresponding to a rest-frame wavelength of $\sim$0.53 \micron). The galaxies are shown when including dust attenuation (top row) and when not including dust attenuation (middle row). The central region of four out of the five example galaxies is much brighter at an observed wavelength of 1.6 \micron in the bottom row than in the top row, i.e., it is brighter when ignoring dust obscuration. Only for the least massive galaxy do we hardly see any change in the brightness of the central region (due to a lack of dust obscuration in low-mass galaxies to begin with, \citealt{Whitaker2017}). The difference in profile becomes clearer in the bottom row of Figure \ref{fig:HbandPoststamps}. Here we show the 1.6 \micron radial light profile of the example galaxies when including dust attenuation and when not including dust attenuation. For all example galaxies, the shape of the profiles is flatter when including dust attenuation. The change in central brightness naturally results in a different half-light radius for these objects when including dust absorption versus when not including dust absorption.

In Figure \ref{fig:HbandRatio} we show the ratio between the 1.6 \micron observed-frame half-light radius of galaxies and the stellar half-mass radius when including dust attenuation (left panel) and when not including dust attenuation (central panel). For the former scenario, we find that the 1.6 \micron half-light radius is more extended than the stellar half-mass radius. The ratio increases as a function of redshift and stellar mass. For the latter scenario (no dust attenuation) the model predicts a ratio between the 1.6 \micron half-light radius and the stellar half-mass radius around unity, independent of stellar mass and redshift. Indeed, the ratio between the two scenarios (right panel of Figure \ref{fig:HbandRatio}) is also larger than one and increases with stellar mass and redshift. The redshift trend is there because the observed-frame 1.6 \micron emission probes shorter rest-frame wavelengths that are more sensitive to dust obscuration with increasing redshift. Future observations focusing on the dust-continuum size of galaxies compared to the H-band size at various redshifts may actually directly test this prediction.

The predictions presented in Figures \ref{fig:HbandPoststamps} and \ref{fig:HbandRatio} suggest the following scenario. The presence of dust attenuates the central bright 1.6 \micron emission of galaxies. Because the central bright component is attenuated, the half-light radius of the galaxies becomes larger compared to the scenario where no dust attenuation takes place. Thus, although the unobscured 1.6 \micron emission very closely follows the stellar distribution, dust attenuation skews the light profile. This becomes increasingly more important at higher redshifts, since the 1.6 \micron emission traces rest-frame wavelengths that are increasingly more susceptible to dust attenuation. The dust emission on the other hand dominates at the locations where the attenuation is strongest (i.e., most light is reprocessed). This is also visible in Figure \ref{fig:poststamps}, where the dust-continuum emission at rest-frame 300 and 850 \micron is more centrally concentrated than the emission at UV wavelengths.

The above suggested scenario provides a natural explanation to why the observed dust emission appears more compact than the observed NIR emission of galaxies. Dust obscuration increases the observed optical/NIR half-light radius compared to the distribution of the dust-continuum emission. Our results suggest that the compact dust distribution is not necessarily a sign of the buildup of a dense central stellar component. On the contrary, on average the predicted dust-continuum half-light radius is similar to the stellar half-mass radius, indicating that the dust-obscured star formation follows the stellar mass distribution relatively closely. This is a common feature of galaxies on the main-sequence and not a peculiarity belonging to a sub-class of main-sequence galaxies. This conclusion does not depend strongly on the underlying choices made for the radiative transfer calculations (e.g., the inclusion of birth clouds).

The presented conclusion has implications for future observations of the resolved dust-continuum distribution of galaxies and their interpretation. When planning ALMA observations aiming at resolving the dust-continuum emission of galaxies, one should expect the galaxies to be smaller than inferred from optical/NIR observations (about 25\% at $z=2$ and more than 50\% at $z=5$). This means that to resolve the dust-continuum emission of galaxies at $z>1$, more extended ALMA configurations may be necessary than one would conclude from the optical/NIR morphology.

\subsection{Single-band dust-continuum emission as a tracer of the dust and star formation distribution in galaxies?}
\label{sec:discussion_dustcontinuumorigin}
In this work we have explored the ratio between the dust-continuum half-light radius of galaxies and various other radii, including the \h2, dust and stellar half-mass radii, and the radius containing half the SFR. Of these, we found that the ratio with the radius containing half the SFR of galaxies is closest to unity. The ratio with the dust half-mass radius on the other hand is typically 0.75.

These results may appear to be surprising, as one may expect that the dust-continuum emission traces the dust distribution in galaxies. To zeroth order, the dust emission of galaxies depends on the temperature of the dust, as well as the amount of dust available (for simplicity ignoring the properties of the dust itself). The temperature dependence is either linear for wavelengths that fall in the Rayleigh-Jeans regime, or even stronger at shorter wavelengths. The temperature of the dust depends on the strength of the UV radiation field from young stars impinging on the dust grains. 

The ratio of roughly unity between the dust-continuum emission half-light radius and the radius containing half the SFR (versus a ratio of 0.75 with the dust half-mass radius) suggests that the dependence on dust temperature is in practice more important than the dependence on dust mass. We checked that the same is true for observed-wavelengths shortward (at least down to 150 \micron) and longward of 850 \micron. We thus conclude that the single-band dust-emission is a decent tracer of the distribution of obscured star formation rate in galaxies and a less robust tracer of the distribution of dust itself (independent of observed wavelength running from 100 \micron to 3 mm).

\subsection{Does the single-band dust-continuum emission of galaxies trace the \h2 distribution?}
In recent years, the integrated single-band dust-continuum emission of galaxies has often been used as a tracer of the \h2 mass of galaxies \citep[e.g.,][]{Scoville2013, Scoville2014,Scoville2016, Scoville2017,Eales2012, Bourne2013, Hughes2017, Schinnerer2016,Tacconi2018}. In this work (Figure \ref{fig:H2Ratio}) we showed that our model predicts ratios between the 850 \micron observed-frame half-light size and \h2 half-mass size of galaxies of 0.75 and below for massive galaxies ($M_{\rm star} > 10^{10}\,\rm{M}_\odot$) at $z \geq 2$ (independent of the adopted \h2 recipe). At these redshifts galaxies with these masses are typically the ones  most easily to resolve by ALMA due to their brightness. The low ratio suggests that the resolved 850 \micron dust-continuum emission of galaxies traces only the central component of the \h2 mass distribution at $z>2$.

Observations also found that the dust-continuum emission is more compact than the CO emission tracing \h2 \citep{Hodge2015, Chen2017,  CalistroRivera2018,Kaasinen2020}. Even though we do not provide predictions for CO emission directly, our predictions suggest that this is true for the majority of main-sequence galaxies and that the dust becomes increasingly more compact compared to \h2 with increasing redshift. We have demonstrated in the right-hand panel of Figure \ref{fig:SFRDustRatio} that the dust-continuum emission closely follows the radius containing half of the star formation. The UV photons from the star forming regions heat up the dust and allow the dust to efficiently emit in the sub-mm wavelength regime. Thus, in our model the underlying driver of the ratio between 850 \micron and \h2 mass is actually the ratio in the size of the star forming region and the \h2 mass.  We checked that this ratio indeed closely follows the trend shown in Figure \ref{fig:H2Ratio}. 

In the presented simulation the SFR of a gas cell is not coupled to its \h2 mass (since the \h2 mass is calculated in post-processing). Therefore, there does not need to be a linear relation between the \h2 mass of a cell and its SFR. It is thus possible for a cell to contain molecular hydrogen, with only little or no star formation at all (see \citealt{Diemer2018} and \citealt{Popping2019}). This naturally allows for a more extended \h2 disk than SFR or dust-continuum disk. But more importantly, dust strongly emits radiation at locations where the dust is heated by emission from young stars (resulting in high dust temperatures). These do not necessarily overlap with the locations where the \h2 mass and CO emission dominate. In a future work we will focus in more detail on the highly resolved emission properties of galaxies.

\section{Summary and Conclusions}
\label{sec:Conclusions}
In this work we have presented predictions for the dust-continuum half-light radius of simulated galaxies, obtained by coupling the radiative transfer code \texttt{SKIRT} to the output of the TNG50 simulation and by adopting as our fiducial ansatz a Milky Way dust type and a metallicity dependent dust-to-metal ratio (based on \citealt{RemyRuyer2014}). The selected simulated galaxies all lie near the star forming main-sequence, have stellar masses between $10^{9}$ and $10^{11}\,\rm{M}_\odot$ and fall in the redshift range $1<z<5$. We compared the dust-continuum half-light radii of the simulated galaxies (with special focus on observed-frame 850\micron) with their stellar, dust and \h2 half-mass radii and the radius containing half the star formation. Our results pertaining to TNG50 galaxies can be summarized as follows.
\begin{itemize}

\item The dust-continuum size of a galaxy is roughly constant (within $\sim 5-10$\%) as a function of wavelength running from 700 \micron to 1 mm (corresponding to ALMA bands 6 though 8). The sizes are within $\sim20$\% of each other in the wavelength regime going from 500 \micron to 2 mm (roughly corresponding to ALMA bands 4 though 9, see Figure \ref{fig:wavelength}).\\

\item The predicted 850 \micron half-light radius of galaxies increases as a function of cosmic time and is constant (shows a very mild increase) as a function of stellar mass for galaxies at $z\geq2$ ($z=1$, see Figure \ref{fig:size_evolution}).\\

\item The predicted observed-frame 850 \micron half-light radius is smaller than the observed-frame 1.6 \micron half-light radius (corresponding to the H-band size), as also seen in observations (Figure \ref{fig:StellarRatio}). The ratio between the half-light radii decreases as a function of redshift, from 0.8 at $z=1$ to less than 0.6 at $z\geq2$ for galaxies more massive than $10^{10}\,\rm{M}_\odot$. This ratio is driven by the obscuration of the 1.6 \micron emission, increasing the observed 1.6 \micron half-light radius of galaxies (Figure \ref{fig:HbandRatio}). The relative compactness increases with redshift, because the observed 1.6 \micron emission corresponds to increasingly bluer rest-frame wavelengths that are more affected by dust attenuation as a function of redshift.\\

\item On the other hand, the 850 \micron half-light radius is typically larger (similar) than the stellar half-mass radius at $z\leq 3$ ($z \geq 4$; Figure~\ref{fig:StellarRatio}). \\

\item The previous points are a common feature of galaxies on the main-sequence of star formation. Combined, they suggest that the relatively compact dust distribution seen in observations is not necessarily a sign of the  dust-obscured buildup of a central dense stellar component, rather a reflection of the change of the stellar light profile of galaxies by dust.\\

\item The 850 \micron half-light radius of galaxies is almost identical to the radius containing half the star formation, to better than 15 per cent in the medians throughout the $10^{9-11}\,\rm{M}_\odot$ and $1<z<5$ studied mass ranges, suggesting that the dust-continuum emission is very closely related to the ongoing star formation in galaxies (Figure \ref{fig:SFRDustRatio}).\\

\item The 850 \micron half-light radius is not necessarily a good tracer of the resolved \h2 size of galaxies at $z\geq2$ (Figure \ref{fig:H2Ratio}). For galaxies with stellar masses larger than $10^{10}\,\rm{M}_\odot$ the dust-continuum size is 0.75 times smaller than the \h2 half-mass radius at $z=2$, and this fraction further decreases towards higher redshifts. The dust emission strongly correlates with locations with the highest dust temperatures, which are not necessarily the locations where most \h2 is located. \\
    
\item The 850 \micron half-light radius of galaxies is about 25 per cent smaller than the dust half-mass radius throughout the studied ranges of  $10^{9-11}\,\rm{M}_\odot$ and $1<z<5$ in mass and redshift, respectively (Figure \ref{fig:SFRDustRatio}). This indicates that the dust emission is not a robust tracer of the underlying dust mass, driven by the dust emission dominating in regions with high dust temperatures.
\end{itemize}

\subsection{Implications for (future) observations}
The presented results have a number of implications for (the interpretation of) future observations: 
\begin{enumerate}
    \item Efforts to resolve the dust-continuum emission of large numbers of galaxies will require ALMA antenna configurations that resolve the galaxy at scales significantly smaller than their optical/NIR size. \\
    \item The use of the dust-continuum emission of galaxies as a reliable tracer of the resolved \h2 and dust mass distribution will require multi-band coverage of the dust SED to account for changes in the dust temperature (and thus the strength of dust emission) across the galaxy disks.\\
    \item The single band dust-continuum emission can be used as a decent tracer of the locations where dust-obscured star formation dominates.\\
    \item The dust-continuum sizes of galaxies at $1<z<5$ as determined from ALMA observations using bands 4 through 7 can safely be compared to each other.
\end{enumerate}

The results presented in this work can serve as a guidance for future observations of the dust continuum in galaxies. We believe this is particularly timely given current and ongoing efforts with ALMA. We thus look forward to future observations further constraining our model predictions for the resolved dust-continuum properties of main-sequence galaxies, pushing forward the connection between gas and star formation in galaxies over cosmic time.

\section*{Acknowledgements}
GP thanks Lichen Liang, Christoph Behrens, Roberto Decarli, Rob Ivison, Gandhali Joshi, Georgios Magdis and Fabian Walter for useful discussions. Simulations for this work were performed on the Isaac supercomputer at the Max Planck Computing and Data Facility, while the TNG50 flagship run was realised with compute time granted by the Gauss Centre for Super-computing (GCS) under GCS Large-Scale Project GCS-DWAR (2016; PIs Nelson/Pillepich). MK acknowledges support from the International Max Planck Research School for Astronomy and Cosmic Physics at Heidelberg University (IMPRS-HD). FM acknowledges support through the Program "Rita Levi Montalcini" of the Italian MUR. We acknowledge use of the \texttt{python} programming language version 3.7, available at \url{http://www.python.org}, Astropy \citep{astropy:2013,astropy:2018}, Matplotlib \citep{Hunter2007}, NumPy \citep{oliphant2006guide,van2011numpy}, SciPy \citep{Virtanen2020SciPy-NMeth} and Seaborn \citep{seaborn}.

\section*{Data availability}
The data that support the findings of this study are available on request from the corresponding author. All data pertaining to the IllustrisTNG project, including the TNG50 run analyzed here, is already openly available \citep{Nelson2019Release} and can be retrieved from the IllustrisTNG website, at \url{www.tng-project.org/data}.




\bibliographystyle{mnras}
\bibliography{DustContinuumSizes} 

\begin{thebibliography}{}
\makeatletter
\relax
\def\mn@urlcharsother{\let\do\@makeother \do\$\do\&\do\#\do\^\do\_\do\%\do\~}
\def\mn@doi{\begingroup\mn@urlcharsother \@ifnextchar [ {\mn@doi@}
  {\mn@doi@[]}}
\def\mn@doi@[#1]#2{\def\@tempa{#1}\ifx\@tempa\@empty \href
  {http://dx.doi.org/#2} {doi:#2}\else \href {http://dx.doi.org/#2} {#1}\fi
  \endgroup}
\def\mn@eprint#1#2{\mn@eprint@#1:#2::\@nil}
\def\mn@eprint@arXiv#1{\href {http://arxiv.org/abs/#1} {{\tt arXiv:#1}}}
\def\mn@eprint@dblp#1{\href {http://dblp.uni-trier.de/rec/bibtex/#1.xml}
  {dblp:#1}}
\def\mn@eprint@#1:#2:#3:#4\@nil{\def\@tempa {#1}\def\@tempb {#2}\def\@tempc
  {#3}\ifx \@tempc \@empty \let \@tempc \@tempb \let \@tempb \@tempa \fi \ifx
  \@tempb \@empty \def\@tempb {arXiv}\fi \@ifundefined
  {mn@eprint@\@tempb}{\@tempb:\@tempc}{\expandafter \expandafter \csname
  mn@eprint@\@tempb\endcsname \expandafter{\@tempc}}}

\bibitem[\protect\citeauthoryear{{Aravena} et~al.,}{{Aravena}
  et~al.}{2020}]{Aravena2020}
{Aravena} M.,  et~al., 2020, arXiv e-prints, \href
  {https://ui.adsabs.harvard.edu/abs/2020arXiv200604284A} {p. arXiv:2006.04284}

\bibitem[\protect\citeauthoryear{{Astropy Collaboration} et~al.,}{{Astropy
  Collaboration} et~al.}{2013}]{astropy:2013}
{Astropy Collaboration} et~al., 2013, \mn@doi [\aap]
  {10.1051/0004-6361/201322068}, \href
  {http://adsabs.harvard.edu/abs/2013A%26A...558A..33A} {558, A33}

\bibitem[\protect\citeauthoryear{{Baes} \& {Camps}}{{Baes} \&
  {Camps}}{2015}]{Baes2015}
{Baes} M.,  {Camps} P.,  2015, \mn@doi [Astronomy and Computing]
  {10.1016/j.ascom.2015.05.006}, \href
  {https://ui.adsabs.harvard.edu/abs/2015A&C....12...33B} {12, 33}

\bibitem[\protect\citeauthoryear{{Baes}, {Verstappen}, {De Looze}, {Fritz},
  {Saftly}, {Vidal P{\'e}rez}, {Stalevski}  \& {Valcke}}{{Baes}
  et~al.}{2011}]{Baes2011}
{Baes} M.,  {Verstappen} J.,  {De Looze} I.,  {Fritz} J.,  {Saftly} W.,  {Vidal
  P{\'e}rez} E.,  {Stalevski} M.,   {Valcke} S.,  2011, \mn@doi [\apjs]
  {10.1088/0067-0049/196/2/22}, \href
  {https://ui.adsabs.harvard.edu/abs/2011ApJS..196...22B} {196, 22}

\bibitem[\protect\citeauthoryear{{Barro} et~al.,}{{Barro}
  et~al.}{2016}]{Barro2016}
{Barro} G.,  et~al., 2016, \mn@doi [\apjl] {10.3847/2041-8205/827/2/L32}, \href
  {https://ui.adsabs.harvard.edu/abs/2016ApJ...827L..32B} {827, L32}

\bibitem[\protect\citeauthoryear{{Behrens}, {Pallottini}, {Ferrara},
  {Gallerani}  \& {Vallini}}{{Behrens} et~al.}{2018}]{Behrens2018}
{Behrens} C.,  {Pallottini} A.,  {Ferrara} A.,  {Gallerani} S.,   {Vallini} L.,
   2018, \mn@doi [\mnras] {10.1093/mnras/sty552}, \href
  {https://ui.adsabs.harvard.edu/abs/2018MNRAS.477..552B} {477, 552}

\bibitem[\protect\citeauthoryear{{Bertemes} et~al.,}{{Bertemes}
  et~al.}{2018}]{Bertemes2018}
{Bertemes} C.,  et~al., 2018, \mn@doi [\mnras] {10.1093/mnras/sty963}, \href
  {https://ui.adsabs.harvard.edu/abs/2018MNRAS.478.1442B} {478, 1442}

\bibitem[\protect\citeauthoryear{{Blitz} \& {Rosolowsky}}{{Blitz} \&
  {Rosolowsky}}{2006}]{Blitz2006}
{Blitz} L.,  {Rosolowsky} E.,  2006, \mn@doi [\apj] {10.1086/505417}, \href
  {http://adsabs.harvard.edu/abs/2006ApJ...650..933B} {650, 933}

\bibitem[\protect\citeauthoryear{{Bourne} et~al.,}{{Bourne}
  et~al.}{2013}]{Bourne2013}
{Bourne} N.,  et~al., 2013, \mn@doi [\mnras] {10.1093/mnras/stt1584}, \href
  {https://ui.adsabs.harvard.edu/abs/2013MNRAS.436..479B} {436, 479}

\bibitem[\protect\citeauthoryear{{Bruzual} \& {Charlot}}{{Bruzual} \&
  {Charlot}}{2003}]{Bruzual2003}
{Bruzual} G.,  {Charlot} S.,  2003, \mn@doi [\mnras]
  {10.1046/j.1365-8711.2003.06897.x}, \href
  {https://ui.adsabs.harvard.edu/abs/2003MNRAS.344.1000B} {344, 1000}

\bibitem[\protect\citeauthoryear{{Calistro Rivera} et~al.,}{{Calistro Rivera}
  et~al.}{2018}]{CalistroRivera2018}
{Calistro Rivera} G.,  et~al., 2018, \mn@doi [\apj] {10.3847/1538-4357/aacffa},
  \href {https://ui.adsabs.harvard.edu/abs/2018ApJ...863...56C} {863, 56}

\bibitem[\protect\citeauthoryear{{Camps} \& {Baes}}{{Camps} \&
  {Baes}}{2015}]{Camps2015}
{Camps} P.,  {Baes} M.,  2015, \mn@doi [Astronomy and Computing]
  {10.1016/j.ascom.2014.10.004}, \href
  {https://ui.adsabs.harvard.edu/abs/2015A&C.....9...20C} {9, 20}

\bibitem[\protect\citeauthoryear{{Camps}, {Baes}  \& {Saftly}}{{Camps}
  et~al.}{2013}]{Camps2013}
{Camps} P.,  {Baes} M.,   {Saftly} W.,  2013, \mn@doi [\aap]
  {10.1051/0004-6361/201322281}, \href
  {https://ui.adsabs.harvard.edu/abs/2013A&A...560A..35C} {560, A35}

\bibitem[\protect\citeauthoryear{{Camps}, {Trayford}, {Baes}, {Theuns},
  {Schaller}  \& {Schaye}}{{Camps} et~al.}{2016}]{Camps2016}
{Camps} P.,  {Trayford} J.~W.,  {Baes} M.,  {Theuns} T.,  {Schaller} M.,
  {Schaye} J.,  2016, \mn@doi [\mnras] {10.1093/mnras/stw1735}, \href
  {https://ui.adsabs.harvard.edu/abs/2016MNRAS.462.1057C} {462, 1057}

\bibitem[\protect\citeauthoryear{{Chen} et~al.,}{{Chen}
  et~al.}{2017}]{Chen2017}
{Chen} C.-C.,  et~al., 2017, \mn@doi [\apj] {10.3847/1538-4357/aa863a}, \href
  {https://ui.adsabs.harvard.edu/abs/2017ApJ...846..108C} {846, 108}

\bibitem[\protect\citeauthoryear{{Cochrane} et~al.,}{{Cochrane}
  et~al.}{2019}]{Cochrane2019}
{Cochrane} R.~K.,  et~al., 2019, \mn@doi [\mnras] {10.1093/mnras/stz1736},
  \href {https://ui.adsabs.harvard.edu/abs/2019MNRAS.488.1779C} {488, 1779}

\bibitem[\protect\citeauthoryear{{Daddi}, {Dannerbauer}, {Elbaz}, {Dickinson},
  {Morrison}, {Stern}  \& {Ravindranath}}{{Daddi} et~al.}{2008}]{Daddi2008}
{Daddi} E.,  {Dannerbauer} H.,  {Elbaz} D.,  {Dickinson} M.,  {Morrison} G.,
  {Stern} D.,   {Ravindranath} S.,  2008, \mn@doi [\apjl] {10.1086/527377},
  \href {https://ui.adsabs.harvard.edu/abs/2008ApJ...673L..21D} {673, L21}

\bibitem[\protect\citeauthoryear{{De Cia}, {Ledoux}, {Mattsson}, {Petitjean},
  {Srianand}, {Gavignaud}  \& {Jenkins}}{{De Cia} et~al.}{2016}]{DeCia2016}
{De Cia} A.,  {Ledoux} C.,  {Mattsson} L.,  {Petitjean} P.,  {Srianand} R.,
  {Gavignaud} I.,   {Jenkins} E.~B.,  2016, \mn@doi [\aap]
  {10.1051/0004-6361/201527895}, \href
  {https://ui.adsabs.harvard.edu/abs/2016A&A...596A..97D} {596, A97}

\bibitem[\protect\citeauthoryear{{De Vis} et~al.,}{{De Vis}
  et~al.}{2019}]{DeVis2019}
{De Vis} P.,  et~al., 2019, \mn@doi [\aap] {10.1051/0004-6361/201834444}, \href
  {https://ui.adsabs.harvard.edu/abs/2019A&A...623A...5D} {623, A5}

\bibitem[\protect\citeauthoryear{{Diemer} et~al.,}{{Diemer}
  et~al.}{2018}]{Diemer2018}
{Diemer} B.,  et~al., 2018, \mn@doi [\apjs] {10.3847/1538-4365/aae387}, \href
  {https://ui.adsabs.harvard.edu/abs/2018ApJS..238...33D} {238, 33}

\bibitem[\protect\citeauthoryear{{Eales} et~al.,}{{Eales}
  et~al.}{2012}]{Eales2012}
{Eales} S.,  et~al., 2012, \mn@doi [\apj] {10.1088/0004-637X/761/2/168}, \href
  {https://ui.adsabs.harvard.edu/abs/2012ApJ...761..168E} {761, 168}

\bibitem[\protect\citeauthoryear{{Fujimoto}, {Ouchi}, {Shibuya}  \&
  {Nagai}}{{Fujimoto} et~al.}{2017}]{Fujimoto2017}
{Fujimoto} S.,  {Ouchi} M.,  {Shibuya} T.,   {Nagai} H.,  2017, \mn@doi [\apj]
  {10.3847/1538-4357/aa93e6}, \href
  {https://ui.adsabs.harvard.edu/abs/2017ApJ...850...83F} {850, 83}

\bibitem[\protect\citeauthoryear{{Gnedin} \& {Kravtsov}}{{Gnedin} \&
  {Kravtsov}}{2011}]{Gnedin2011}
{Gnedin} N.~Y.,  {Kravtsov} A.~V.,  2011, \mn@doi [\apj]
  {10.1088/0004-637X/728/2/88}, \href
  {https://ui.adsabs.harvard.edu/abs/2011ApJ...728...88G} {728, 88}

\bibitem[\protect\citeauthoryear{{Groves}, {Dopita}, {Sutherland}, {Kewley},
  {Fischera}, {Leitherer}, {Brandl}  \& {van Breugel}}{{Groves}
  et~al.}{2008}]{Groves2008}
{Groves} B.,  {Dopita} M.~A.,  {Sutherland} R.~S.,  {Kewley} L.~J.,  {Fischera}
  J.,  {Leitherer} C.,  {Brandl} B.,   {van Breugel} W.,  2008, \mn@doi [\apjs]
  {10.1086/528711}, \href
  {https://ui.adsabs.harvard.edu/abs/2008ApJS..176..438G} {176, 438}

\bibitem[\protect\citeauthoryear{{Groves} et~al.,}{{Groves}
  et~al.}{2015}]{Groves2015}
{Groves} B.~A.,  et~al., 2015, \mn@doi [\apj] {10.1088/0004-637X/799/1/96},
  \href {https://ui.adsabs.harvard.edu/abs/2015ApJ...799...96G} {799, 96}

\bibitem[\protect\citeauthoryear{{Gullberg} et~al.,}{{Gullberg}
  et~al.}{2019}]{Gullberg2019}
{Gullberg} B.,  et~al., 2019, \mn@doi [\mnras] {10.1093/mnras/stz2835}, \href
  {https://ui.adsabs.harvard.edu/abs/2019MNRAS.490.4956G} {490, 4956}

\bibitem[\protect\citeauthoryear{{Hemler} et~al.,}{{Hemler}
  et~al.}{2020}]{Hemler2020}
{Hemler} Z.~S.,  et~al., 2020, arXiv e-prints, \href
  {https://ui.adsabs.harvard.edu/abs/2020arXiv200710993H} {p. arXiv:2007.10993}

\bibitem[\protect\citeauthoryear{{Hodge} \& {da Cunha}}{{Hodge} \& {da
  Cunha}}{2020}]{Hodge2020}
{Hodge} J.~A.,  {da Cunha} E.,  2020, arXiv e-prints, \href
  {https://ui.adsabs.harvard.edu/abs/2020arXiv200400934H} {p. arXiv:2004.00934}

\bibitem[\protect\citeauthoryear{{Hodge} et~al.,}{{Hodge}
  et~al.}{2013}]{Hodge2013}
{Hodge} J.~A.,  et~al., 2013, \mn@doi [\apj] {10.1088/0004-637X/768/1/91},
  \href {https://ui.adsabs.harvard.edu/abs/2013ApJ...768...91H} {768, 91}

\bibitem[\protect\citeauthoryear{{Hodge}, {Riechers}, {Decarli}, {Walter},
  {Carilli}, {Daddi}  \& {Dannerbauer}}{{Hodge} et~al.}{2015}]{Hodge2015}
{Hodge} J.~A.,  {Riechers} D.,  {Decarli} R.,  {Walter} F.,  {Carilli} C.~L.,
  {Daddi} E.,   {Dannerbauer} H.,  2015, \mn@doi [\apjl]
  {10.1088/2041-8205/798/1/L18}, \href
  {https://ui.adsabs.harvard.edu/abs/2015ApJ...798L..18H} {798, L18}

\bibitem[\protect\citeauthoryear{{Hodge} et~al.,}{{Hodge}
  et~al.}{2019}]{Hodge2019}
{Hodge} J.~A.,  et~al., 2019, \mn@doi [\apj] {10.3847/1538-4357/ab1846}, \href
  {https://ui.adsabs.harvard.edu/abs/2019ApJ...876..130H} {876, 130}

\bibitem[\protect\citeauthoryear{{Hopkins} et~al.,}{{Hopkins}
  et~al.}{2018}]{Hopkins2018}
{Hopkins} P.~F.,  et~al., 2018, \mn@doi [\mnras] {10.1093/mnras/sty1690}, \href
  {https://ui.adsabs.harvard.edu/abs/2018MNRAS.480..800H} {480, 800}

\bibitem[\protect\citeauthoryear{{Hou}, {Aoyama}, {Hirashita}, {Nagamine}  \&
  {Shimizu}}{{Hou} et~al.}{2019}]{Hou2019}
{Hou} K.-C.,  {Aoyama} S.,  {Hirashita} H.,  {Nagamine} K.,   {Shimizu} I.,
  2019, \mn@doi [\mnras] {10.1093/mnras/stz121}, \href
  {https://ui.adsabs.harvard.edu/abs/2019MNRAS.485.1727H} {485, 1727}

\bibitem[\protect\citeauthoryear{{Hughes} et~al.,}{{Hughes}
  et~al.}{2017}]{Hughes2017}
{Hughes} T.~M.,  et~al., 2017, \mn@doi [\mnras] {10.1093/mnrasl/slx033}, \href
  {https://ui.adsabs.harvard.edu/abs/2017MNRAS.468L.103H} {468, L103}

\bibitem[\protect\citeauthoryear{Hunter}{Hunter}{2007}]{Hunter2007}
Hunter J.~D.,  2007, \mn@doi [Computing in Science \& Engineering]
  {10.1109/MCSE.2007.55}, 9, 90

\bibitem[\protect\citeauthoryear{{Jonsson}}{{Jonsson}}{2006}]{Jonsson2006}
{Jonsson} P.,  2006, \mn@doi [\mnras] {10.1111/j.1365-2966.2006.10884.x}, \href
  {https://ui.adsabs.harvard.edu/abs/2006MNRAS.372....2J} {372, 2}

\bibitem[\protect\citeauthoryear{{Kaasinen} et~al.,}{{Kaasinen}
  et~al.}{2019}]{Kaasinen2019}
{Kaasinen} M.,  et~al., 2019, \mn@doi [\apj] {10.3847/1538-4357/ab253b}, \href
  {https://ui.adsabs.harvard.edu/abs/2019ApJ...880...15K} {880, 15}

\bibitem[\protect\citeauthoryear{{Kaasinen} et~al.,}{{Kaasinen}
  et~al.}{2020}]{Kaasinen2020}
{Kaasinen} M.,  et~al., 2020, \mn@doi [\apj] {10.3847/1538-4357/aba438}, \href
  {https://ui.adsabs.harvard.edu/abs/2020ApJ...899...37K} {899, 37}

\bibitem[\protect\citeauthoryear{{Karim} et~al.,}{{Karim}
  et~al.}{2013}]{Karim2013}
{Karim} A.,  et~al., 2013, \mn@doi [\mnras] {10.1093/mnras/stt196}, \href
  {https://ui.adsabs.harvard.edu/abs/2013MNRAS.432....2K} {432, 2}

\bibitem[\protect\citeauthoryear{{Krumholz}}{{Krumholz}}{2013}]{Krumholz2013}
{Krumholz} M.~R.,  2013, \mn@doi [\mnras] {10.1093/mnras/stt1780}, \href
  {http://adsabs.harvard.edu/abs/2013MNRAS.436.2747K} {436, 2747}

\bibitem[\protect\citeauthoryear{{Lagos} et~al.,}{{Lagos}
  et~al.}{2019}]{Lagos2019}
{Lagos} C. d.~P.,  et~al., 2019, \mn@doi [\mnras] {10.1093/mnras/stz2427},
  \href {https://ui.adsabs.harvard.edu/abs/2019MNRAS.489.4196L} {489, 4196}

\bibitem[\protect\citeauthoryear{{Lang} et~al.,}{{Lang}
  et~al.}{2019}]{Lang2019}
{Lang} P.,  et~al., 2019, \mn@doi [\apj] {10.3847/1538-4357/ab1f77}, \href
  {https://ui.adsabs.harvard.edu/abs/2019ApJ...879...54L} {879, 54}

\bibitem[\protect\citeauthoryear{{Li}, {Narayanan}  \& {Dav{\'e}}}{{Li}
  et~al.}{2019}]{Li2019}
{Li} Q.,  {Narayanan} D.,   {Dav{\'e}} R.,  2019, \mn@doi [\mnras]
  {10.1093/mnras/stz2684}, \href
  {https://ui.adsabs.harvard.edu/abs/2019MNRAS.490.1425L} {490, 1425}

\bibitem[\protect\citeauthoryear{{Liang}, {Feldmann}, {Faucher-Gigu{\`e}re},
  {Kere{\v{s}}}, {Hopkins}, {Hayward}, {Quataert}  \& {Scoville}}{{Liang}
  et~al.}{2018}]{Liang2018}
{Liang} L.,  {Feldmann} R.,  {Faucher-Gigu{\`e}re} C.-A.,  {Kere{\v{s}}} D.,
  {Hopkins} P.~F.,  {Hayward} C.~C.,  {Quataert} E.,   {Scoville} N.~Z.,  2018,
  \mn@doi [\mnras] {10.1093/mnrasl/sly071}, \href
  {https://ui.adsabs.harvard.edu/abs/2018MNRAS.478L..83L} {478, L83}

\bibitem[\protect\citeauthoryear{{Liang} et~al.,}{{Liang}
  et~al.}{2019}]{Liang2019}
{Liang} L.,  et~al., 2019, \mn@doi [\mnras] {10.1093/mnras/stz2134}, \href
  {https://ui.adsabs.harvard.edu/abs/2019MNRAS.489.1397L} {489, 1397}

\bibitem[\protect\citeauthoryear{{Liang}, {Feldmann}, {Hayward}, {Narayanan},
  {{\c{C}}atmabacak}, {Kere{\v{s}}}, {Faucher-Gigu{\`e}re}  \&
  {Hopkins}}{{Liang} et~al.}{2020}]{Liang2020}
{Liang} L.,  {Feldmann} R.,  {Hayward} C.~C.,  {Narayanan} D.,
  {{\c{C}}atmabacak} O.,  {Kere{\v{s}}} D.,  {Faucher-Gigu{\`e}re} C.-A.,
  {Hopkins} P.~F.,  2020, arXiv e-prints, \href
  {https://ui.adsabs.harvard.edu/abs/2020arXiv200913522L} {p. arXiv:2009.13522}

\bibitem[\protect\citeauthoryear{{Liu} et~al.,}{{Liu} et~al.}{2019}]{Liu2019}
{Liu} D.,  et~al., 2019, \mn@doi [\apj] {10.3847/1538-4357/ab578d}, \href
  {https://ui.adsabs.harvard.edu/abs/2019ApJ...887..235L} {887, 235}

\bibitem[\protect\citeauthoryear{{Lovell}, {Geach}, {Dav{\'e}}, {Narayanan}  \&
  {Li}}{{Lovell} et~al.}{2020}]{Lovell2020}
{Lovell} C.~C.,  {Geach} J.~E.,  {Dav{\'e}} R.,  {Narayanan} D.,   {Li} Q.,
  2020, arXiv e-prints, \href
  {https://ui.adsabs.harvard.edu/abs/2020arXiv200615156L} {p. arXiv:2006.15156}

\bibitem[\protect\citeauthoryear{{Magnelli} et~al.,}{{Magnelli}
  et~al.}{2020}]{Magnelli2020}
{Magnelli} B.,  et~al., 2020, \mn@doi [\apj] {10.3847/1538-4357/ab7897}, \href
  {https://ui.adsabs.harvard.edu/abs/2020ApJ...892...66M} {892, 66}

\bibitem[\protect\citeauthoryear{{Marinacci} et~al.,}{{Marinacci}
  et~al.}{2018}]{Marinacci2018}
{Marinacci} F.,  et~al., 2018, \mn@doi [\mnras] {10.1093/mnras/sty2206}, \href
  {https://ui.adsabs.harvard.edu/abs/2018MNRAS.480.5113M} {480, 5113}

\bibitem[\protect\citeauthoryear{{McKinnon}, {Torrey}  \&
  {Vogelsberger}}{{McKinnon} et~al.}{2016}]{McKinnon2016}
{McKinnon} R.,  {Torrey} P.,   {Vogelsberger} M.,  2016, \mn@doi [\mnras]
  {10.1093/mnras/stw253}, \href
  {https://ui.adsabs.harvard.edu/abs/2016MNRAS.457.3775M} {457, 3775}

\bibitem[\protect\citeauthoryear{{Millard}, {Diemer}, {Eales}, {Gomez},
  {Beeston}  \& {Smith}}{{Millard} et~al.}{2021}]{Millard2021}
{Millard} J.~S.,  {Diemer} B.,  {Eales} S.~A.,  {Gomez} H.~L.,  {Beeston} R.,
  {Smith} M. W.~L.,  2021, \mn@doi [\mnras] {10.1093/mnras/staa3207}, \href
  {https://ui.adsabs.harvard.edu/abs/2021MNRAS.500..871M} {500, 871}

\bibitem[\protect\citeauthoryear{{Naiman} et~al.,}{{Naiman}
  et~al.}{2018}]{Naiman2018}
{Naiman} J.~P.,  et~al., 2018, \mn@doi [\mnras] {10.1093/mnras/sty618}, \href
  {https://ui.adsabs.harvard.edu/abs/2018MNRAS.477.1206N} {477, 1206}

\bibitem[\protect\citeauthoryear{{Narayanan} et~al.,}{{Narayanan}
  et~al.}{2020}]{Narayanan2020}
{Narayanan} D.,  et~al., 2020, arXiv e-prints, \href
  {https://ui.adsabs.harvard.edu/abs/2020arXiv200610757N} {p. arXiv:2006.10757}

\bibitem[\protect\citeauthoryear{{Nelson} et~al.,}{{Nelson}
  et~al.}{2018}]{Nelson2018}
{Nelson} D.,  et~al., 2018, \mn@doi [\mnras] {10.1093/mnras/stx3040}, \href
  {https://ui.adsabs.harvard.edu/abs/2018MNRAS.475..624N} {475, 624}

\bibitem[\protect\citeauthoryear{{Nelson} et~al.,}{{Nelson}
  et~al.}{2019a}]{Nelson2019Release}
{Nelson} D.,  et~al., 2019a, \mn@doi [Computational Astrophysics and Cosmology]
  {10.1186/s40668-019-0028-x}, \href
  {https://ui.adsabs.harvard.edu/abs/2019ComAC...6....2N} {6, 2}

\bibitem[\protect\citeauthoryear{{Nelson} et~al.,}{{Nelson}
  et~al.}{2019b}]{Nelson2019}
{Nelson} D.,  et~al., 2019b, \mn@doi [\mnras] {10.1093/mnras/stz2306}, \href
  {https://ui.adsabs.harvard.edu/abs/2019MNRAS.490.3234N} {490, 3234}

\bibitem[\protect\citeauthoryear{{Nelson} et~al.,}{{Nelson}
  et~al.}{2019c}]{EricaNelson2018}
{Nelson} E.~J.,  et~al., 2019c, \mn@doi [\apj] {10.3847/1538-4357/aaf38a},
  \href {https://ui.adsabs.harvard.edu/abs/2019ApJ...870..130N} {870, 130}

\bibitem[\protect\citeauthoryear{{Noeske} et~al.,}{{Noeske}
  et~al.}{2007}]{Noeske2007}
{Noeske} K.~G.,  et~al., 2007, \mn@doi [\apjl] {10.1086/517926}, \href
  {https://ui.adsabs.harvard.edu/abs/2007ApJ...660L..43N} {660, L43}

\bibitem[\protect\citeauthoryear{Oliphant}{Oliphant}{2006}]{oliphant2006guide}
Oliphant T.~E.,  2006, A guide to NumPy.
~"" Vol. 1, Trelgol Publishing USA

\bibitem[\protect\citeauthoryear{{P{\'e}roux} \& {Howk}}{{P{\'e}roux} \&
  {Howk}}{2020}]{Peroux2020}
{P{\'e}roux} C.,  {Howk} J.~C.,  2020, \mn@doi [\araa]
  {10.1146/annurev-astro-021820-120014}, \href
  {https://ui.adsabs.harvard.edu/abs/2020ARA&A..5821820P} {58, annurev}

\bibitem[\protect\citeauthoryear{{Pillepich} et~al.,}{{Pillepich}
  et~al.}{2018a}]{Pillepich2018a}
{Pillepich} A.,  et~al., 2018a, \mn@doi [\mnras] {10.1093/mnras/stx2656}, \href
  {https://ui.adsabs.harvard.edu/abs/2018MNRAS.473.4077P} {473, 4077}

\bibitem[\protect\citeauthoryear{{Pillepich} et~al.,}{{Pillepich}
  et~al.}{2018b}]{Pillepich2018b}
{Pillepich} A.,  et~al., 2018b, \mn@doi [\mnras] {10.1093/mnras/stx3112}, \href
  {https://ui.adsabs.harvard.edu/abs/2018MNRAS.475..648P} {475, 648}

\bibitem[\protect\citeauthoryear{{Pillepich} et~al.,}{{Pillepich}
  et~al.}{2019}]{Pillepich2019}
{Pillepich} A.,  et~al., 2019, \mn@doi [\mnras] {10.1093/mnras/stz2338}, \href
  {https://ui.adsabs.harvard.edu/abs/2019MNRAS.490.3196P} {490, 3196}

\bibitem[\protect\citeauthoryear{{Planck Collaboration} et~al.,}{{Planck
  Collaboration} et~al.}{2016}]{Planck2016}
{Planck Collaboration} et~al., 2016, \mn@doi [\aap]
  {10.1051/0004-6361/201525830}, \href
  {https://ui.adsabs.harvard.edu/abs/2016A&A...594A..13P} {594, A13}

\bibitem[\protect\citeauthoryear{{Popping}, {Somerville}  \&
  {Galametz}}{{Popping} et~al.}{2017a}]{Popping2017dust}
{Popping} G.,  {Somerville} R.~S.,   {Galametz} M.,  2017a, \mn@doi [\mnras]
  {10.1093/mnras/stx1545}, \href
  {https://ui.adsabs.harvard.edu/abs/2017MNRAS.471.3152P} {471, 3152}

\bibitem[\protect\citeauthoryear{{Popping} et~al.,}{{Popping}
  et~al.}{2017b}]{Popping2017}
{Popping} G.,  et~al., 2017b, \mn@doi [\aap] {10.1051/0004-6361/201730391},
  \href {https://ui.adsabs.harvard.edu/abs/2017A&A...602A..11P} {602, A11}

\bibitem[\protect\citeauthoryear{{Popping} et~al.,}{{Popping}
  et~al.}{2019}]{Popping2019}
{Popping} G.,  et~al., 2019, \mn@doi [\apj] {10.3847/1538-4357/ab30f2}, \href
  {https://ui.adsabs.harvard.edu/abs/2019ApJ...882..137P} {882, 137}

\bibitem[\protect\citeauthoryear{{Price-Whelan} et~al.,}{{Price-Whelan}
  et~al.}{2018}]{astropy:2018}
{Price-Whelan} A.~M.,  et~al., 2018, \mn@doi [\aj] {10.3847/1538-3881/aabc4f},
  \href {https://ui.adsabs.harvard.edu/#abs/2018AJ....156..123T} {156, 123}

\bibitem[\protect\citeauthoryear{{Privon}, {Narayanan}  \& {Dav{\'e}}}{{Privon}
  et~al.}{2018}]{Privon2018}
{Privon} G.~C.,  {Narayanan} D.,   {Dav{\'e}} R.,  2018, \mn@doi [\apj]
  {10.3847/1538-4357/aae485}, \href
  {https://ui.adsabs.harvard.edu/abs/2018ApJ...867..102P} {867, 102}

\bibitem[\protect\citeauthoryear{{Puglisi} et~al.,}{{Puglisi}
  et~al.}{2019}]{Puglisi2019}
{Puglisi} A.,  et~al., 2019, \mn@doi [\apjl] {10.3847/2041-8213/ab1f92}, \href
  {https://ui.adsabs.harvard.edu/abs/2019ApJ...877L..23P} {877, L23}

\bibitem[\protect\citeauthoryear{{R{\'e}my-Ruyer} et~al.,}{{R{\'e}my-Ruyer}
  et~al.}{2014}]{RemyRuyer2014}
{R{\'e}my-Ruyer} A.,  et~al., 2014, \mn@doi [\aap]
  {10.1051/0004-6361/201322803}, \href
  {https://ui.adsabs.harvard.edu/abs/2014A&A...563A..31R} {563, A31}

\bibitem[\protect\citeauthoryear{{Rodriguez-Gomez} et~al.,}{{Rodriguez-Gomez}
  et~al.}{2019}]{Rodriguez2019}
{Rodriguez-Gomez} V.,  et~al., 2019, \mn@doi [\mnras] {10.1093/mnras/sty3345},
  \href {https://ui.adsabs.harvard.edu/abs/2019MNRAS.483.4140R} {483, 4140}

\bibitem[\protect\citeauthoryear{{Rujopakarn} et~al.,}{{Rujopakarn}
  et~al.}{2019}]{Rujopakarn2019}
{Rujopakarn} W.,  et~al., 2019, \mn@doi [\apj] {10.3847/1538-4357/ab3791},
  \href {https://ui.adsabs.harvard.edu/abs/2019ApJ...882..107R} {882, 107}

\bibitem[\protect\citeauthoryear{{Saftly}, {Baes}  \& {Camps}}{{Saftly}
  et~al.}{2014}]{Saftly2014}
{Saftly} W.,  {Baes} M.,   {Camps} P.,  2014, \mn@doi [\aap]
  {10.1051/0004-6361/201322593}, \href
  {https://ui.adsabs.harvard.edu/abs/2014A&A...561A..77S} {561, A77}

\bibitem[\protect\citeauthoryear{{Schinnerer} et~al.,}{{Schinnerer}
  et~al.}{2016}]{Schinnerer2016}
{Schinnerer} E.,  et~al., 2016, \mn@doi [\apj] {10.3847/1538-4357/833/1/112},
  \href {https://ui.adsabs.harvard.edu/abs/2016ApJ...833..112S} {833, 112}

\bibitem[\protect\citeauthoryear{{Schulz}, {Popping}, {Pillepich}, {Nelson},
  {Vogelsberger}, {Marinacci}  \& {Hernquist}}{{Schulz}
  et~al.}{2020}]{Schulz2020}
{Schulz} S.,  {Popping} G.,  {Pillepich} A.,  {Nelson} D.,  {Vogelsberger} M.,
  {Marinacci} F.,   {Hernquist} L.,  2020, \mn@doi [\mnras]
  {10.1093/mnras/staa1900}, \href
  {https://ui.adsabs.harvard.edu/abs/2020MNRAS.tmp.2029S} {}

\bibitem[\protect\citeauthoryear{{Scoville} et~al.,}{{Scoville}
  et~al.}{2013}]{Scoville2013}
{Scoville} N.,  et~al., 2013, \mn@doi [\apjs] {10.1088/0067-0049/206/1/3},
  \href {https://ui.adsabs.harvard.edu/abs/2013ApJS..206....3S} {206, 3}

\bibitem[\protect\citeauthoryear{{Scoville} et~al.,}{{Scoville}
  et~al.}{2014}]{Scoville2014}
{Scoville} N.,  et~al., 2014, \mn@doi [\apj] {10.1088/0004-637X/783/2/84},
  \href {https://ui.adsabs.harvard.edu/abs/2014ApJ...783...84S} {783, 84}

\bibitem[\protect\citeauthoryear{{Scoville} et~al.,}{{Scoville}
  et~al.}{2016}]{Scoville2016}
{Scoville} N.,  et~al., 2016, \mn@doi [\apj] {10.3847/0004-637X/820/2/83},
  \href {https://ui.adsabs.harvard.edu/abs/2016ApJ...820...83S} {820, 83}

\bibitem[\protect\citeauthoryear{{Scoville} et~al.,}{{Scoville}
  et~al.}{2017}]{Scoville2017}
{Scoville} N.,  et~al., 2017, \mn@doi [\apj] {10.3847/1538-4357/aa61a0}, \href
  {https://ui.adsabs.harvard.edu/abs/2017ApJ...837..150S} {837, 150}

\bibitem[\protect\citeauthoryear{{Shapley}, {Cullen}, {Dunlop}, {McLure},
  {Kriek}, {Reddy}  \& {Sand ers}}{{Shapley} et~al.}{2020}]{Shapley2020}
{Shapley} A.~E.,  {Cullen} F.,  {Dunlop} J.~S.,  {McLure} R.~J.,  {Kriek} M.,
  {Reddy} N.~A.,   {Sand ers} R.~L.,  2020, arXiv e-prints, \href
  {https://ui.adsabs.harvard.edu/abs/2020arXiv200910091S} {p. arXiv:2009.10091}

\bibitem[\protect\citeauthoryear{{Simpson} et~al.,}{{Simpson}
  et~al.}{2015}]{Simpson2015}
{Simpson} J.~M.,  et~al., 2015, \mn@doi [\apj] {10.1088/0004-637X/799/1/81},
  \href {https://ui.adsabs.harvard.edu/abs/2015ApJ...799...81S} {799, 81}

\bibitem[\protect\citeauthoryear{{Springel}}{{Springel}}{2010}]{Springel2010}
{Springel} V.,  2010, \mn@doi [\mnras] {10.1111/j.1365-2966.2009.15715.x},
  \href {https://ui.adsabs.harvard.edu/abs/2010MNRAS.401..791S} {401, 791}

\bibitem[\protect\citeauthoryear{{Springel} et~al.,}{{Springel}
  et~al.}{2018}]{Springel2018}
{Springel} V.,  et~al., 2018, \mn@doi [\mnras] {10.1093/mnras/stx3304}, \href
  {https://ui.adsabs.harvard.edu/abs/2018MNRAS.475..676S} {475, 676}

\bibitem[\protect\citeauthoryear{{Stach} et~al.,}{{Stach}
  et~al.}{2019}]{Stach2019}
{Stach} S.~M.,  et~al., 2019, \mn@doi [\mnras] {10.1093/mnras/stz1536}, \href
  {https://ui.adsabs.harvard.edu/abs/2019MNRAS.487.4648S} {487, 4648}

\bibitem[\protect\citeauthoryear{{Tacconi} et~al.,}{{Tacconi}
  et~al.}{2018}]{Tacconi2018}
{Tacconi} L.~J.,  et~al., 2018, \mn@doi [\apj] {10.3847/1538-4357/aaa4b4},
  \href {https://ui.adsabs.harvard.edu/abs/2018ApJ...853..179T} {853, 179}

\bibitem[\protect\citeauthoryear{{Tadaki} et~al.,}{{Tadaki}
  et~al.}{2017}]{Tadaki2017}
{Tadaki} K.-i.,  et~al., 2017, \mn@doi [\apj] {10.3847/1538-4357/834/2/135},
  \href {https://ui.adsabs.harvard.edu/abs/2017ApJ...834..135T} {834, 135}

\bibitem[\protect\citeauthoryear{{Tadaki} et~al.,}{{Tadaki}
  et~al.}{2020}]{Tadaki2020}
{Tadaki} K.-i.,  et~al., 2020, arXiv e-prints, \href
  {https://ui.adsabs.harvard.edu/abs/2020arXiv200901976T} {p. arXiv:2009.01976}

\bibitem[\protect\citeauthoryear{{Torrey}, {Vogelsberger}, {Genel}, {Sijacki},
  {Springel}  \& {Hernquist}}{{Torrey} et~al.}{2014}]{Torrey2014}
{Torrey} P.,  {Vogelsberger} M.,  {Genel} S.,  {Sijacki} D.,  {Springel} V.,
  {Hernquist} L.,  2014, \mn@doi [\mnras] {10.1093/mnras/stt2295}, \href
  {https://ui.adsabs.harvard.edu/abs/2014MNRAS.438.1985T} {438, 1985}

\bibitem[\protect\citeauthoryear{Van Der~Walt, Colbert  \& Varoquaux}{Van
  Der~Walt et~al.}{2011}]{van2011numpy}
Van Der~Walt S.,  Colbert S.~C.,   Varoquaux G.,  2011, Computing in Science \&
  Engineering, 13, 22

\bibitem[\protect\citeauthoryear{{Virtanen} et~al.,}{{Virtanen}
  et~al.}{2020}]{Virtanen2020SciPy-NMeth}
{Virtanen} P.,  et~al., 2020, \mn@doi [Nature Methods]
  {https://doi.org/10.1038/s41592-019-0686-2}, \href {https://rdcu.be/b08Wh}
  {17, 261}

\bibitem[\protect\citeauthoryear{{Vogelsberger}, {Genel}, {Sijacki}, {Torrey},
  {Springel}  \& {Hernquist}}{{Vogelsberger} et~al.}{2013}]{Vogelsberger2013}
{Vogelsberger} M.,  {Genel} S.,  {Sijacki} D.,  {Torrey} P.,  {Springel} V.,
  {Hernquist} L.,  2013, \mn@doi [\mnras] {10.1093/mnras/stt1789}, \href
  {https://ui.adsabs.harvard.edu/abs/2013MNRAS.436.3031V} {436, 3031}

\bibitem[\protect\citeauthoryear{{Vogelsberger} et~al.,}{{Vogelsberger}
  et~al.}{2020}]{Vogelsberger2019}
{Vogelsberger} M.,  et~al., 2020, \mn@doi [\mnras] {10.1093/mnras/staa137},
  \href {https://ui.adsabs.harvard.edu/abs/2020MNRAS.492.5167V} {492, 5167}

\bibitem[\protect\citeauthoryear{Waskom et~al.,}{Waskom et~al.}{2017}]{seaborn}
Waskom M.,  et~al., 2017, mwaskom/seaborn: v0.8.1 (September 2017),
  \mn@doi{10.5281/zenodo.883859}, \url {https://doi.org/10.5281/zenodo.883859}

\bibitem[\protect\citeauthoryear{{Weinberger} et~al.,}{{Weinberger}
  et~al.}{2017}]{Weinberger2017}
{Weinberger} R.,  et~al., 2017, \mn@doi [\mnras] {10.1093/mnras/stw2944}, \href
  {https://ui.adsabs.harvard.edu/abs/2017MNRAS.465.3291W} {465, 3291}

\bibitem[\protect\citeauthoryear{{Weingartner} \& {Draine}}{{Weingartner} \&
  {Draine}}{2001}]{Weingartner2001}
{Weingartner} J.~C.,  {Draine} B.~T.,  2001, \mn@doi [\apj] {10.1086/318651},
  \href {https://ui.adsabs.harvard.edu/abs/2001ApJ...548..296W} {548, 296}

\bibitem[\protect\citeauthoryear{{Whitaker} et~al.,}{{Whitaker}
  et~al.}{2014}]{Whitaker2014}
{Whitaker} K.~E.,  et~al., 2014, \mn@doi [\apj] {10.1088/0004-637X/795/2/104},
  \href {https://ui.adsabs.harvard.edu/abs/2014ApJ...795..104W} {795, 104}

\bibitem[\protect\citeauthoryear{{Whitaker}, {Pope}, {Cybulski}, {Casey},
  {Popping}  \& {Yun}}{{Whitaker} et~al.}{2017}]{Whitaker2017}
{Whitaker} K.~E.,  {Pope} A.,  {Cybulski} R.,  {Casey} C.~M.,  {Popping} G.,
  {Yun} M.~S.,  2017, \mn@doi [\apj] {10.3847/1538-4357/aa94ce}, \href
  {https://ui.adsabs.harvard.edu/abs/2017ApJ...850..208W} {850, 208}

\bibitem[\protect\citeauthoryear{{Wiseman}, {Schady}, {Bolmer}, {Kr{\"u}hler},
  {Yates}, {Greiner}  \& {Fynbo}}{{Wiseman} et~al.}{2017}]{Wiseman2017}
{Wiseman} P.,  {Schady} P.,  {Bolmer} J.,  {Kr{\"u}hler} T.,  {Yates} R.~M.,
  {Greiner} J.,   {Fynbo} J.~P.~U.,  2017, \mn@doi [\aap]
  {10.1051/0004-6361/201629228}, \href
  {https://ui.adsabs.harvard.edu/abs/2017A&A...599A..24W} {599, A24}

\bibitem[\protect\citeauthoryear{{da Cunha} et~al.,}{{da Cunha}
  et~al.}{2013}]{daCunha2013}
{da Cunha} E.,  et~al., 2013, \mn@doi [\apj] {10.1088/0004-637X/766/1/13},
  \href {https://ui.adsabs.harvard.edu/abs/2013ApJ...766...13D} {766, 13}

\bibitem[\protect\citeauthoryear{{van der Wel} et~al.,}{{van der Wel}
  et~al.}{2014}]{vanderWel2013}
{van der Wel} A.,  et~al., 2014, \mn@doi [\apj] {10.1088/0004-637X/788/1/28},
  \href {https://ui.adsabs.harvard.edu/abs/2014ApJ...788...28V} {788, 28}

\makeatother
\end{thebibliography}

\appendix
\section{Dependence of the 1.6 and 850 \micron half-light radius on the dust-to-metal ratio}
\label{sec:appendixDTM}
Our fiducial model adopts a dust-to-metal mass ratio (DTM) that varies as a function of metallicity, with DTM increasing for increasing metallicity, up to a value of 0.32 at a metallicity of 0.4 times solar. This is motivated by observations of dust and CO (and HI) emission in local (and a handful of $z\sim2$) galaxies (\citealt{RemyRuyer2014}, \citealt{DeVis2019},\citealt{Shapley2020}, see Figure 1 in \citealt{Lagos2019}) and high-redshift damped Lyman-alpha and gamma-ray absorbers \citep[e.g.,][]{Peroux2020}.  Simulations that track the production and destruction of dust have also found the DTM to vary with galaxy properties \citep[e.g.,][]{McKinnon2016, Popping2017, Li2019}. Other simulations found that they require a variable DTM in order to reproduce the UV luminosity of low-mass high-redshift galaxies \citep{Lagos2019,Vogelsberger2019}.
This assumption is different from the classical one of a constant DTM of $\sim 0.3$ often adopted in simulation-based works \citep[see for example][]{Cochrane2019, Schulz2020, Liang2020, Millard2021}.

\begin{figure*}
\begin{center}
\includegraphics[width = 1.0\hsize]{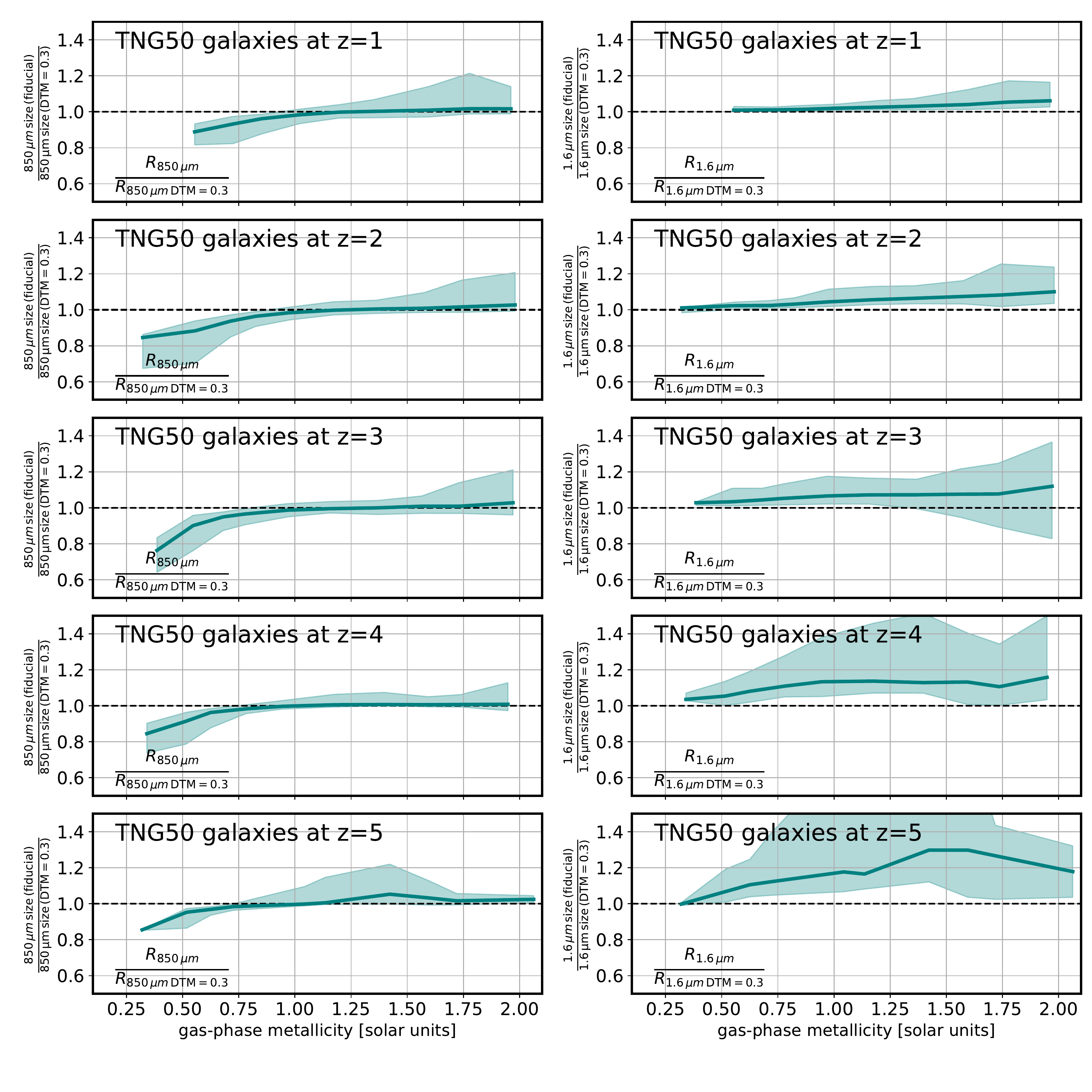}
\caption{Left column: The ratio between the 850 \micron half-light radius of the TNG50+\texttt{SKIRT} model galaxies when adopting our fiducial metallicity-dependent dust-to-metal mass ratio (DTM) and the half-light radius obtained when adopting a fixed DTM of 0.3, as a function of the mass-weighted gas-phase metallicity of the galaxies. Right column: The same as the left column but for the 1.6 \micron half-light radius. The rows indicate different redshifts. When adopting a fixed DTM of 0.3 the model predicts more extended 850 \micron emission and more compact 1.6 \micron emission compared to our fiducial choice for the DTM. Differences are all within 10-20 per cent from our fiducial model. \label{fig:dtm_variation}}
\end{center}
\end{figure*}

To estimate the importance of the choice of DTM on our results, we re-ran the radiative transfer code \texttt{SKIRT} on all the modeled galaxies assuming a constant and universal DTM of 0.3. In Figure \ref{fig:dtm_variation} we show the ratio between the 850 and 1.6 \micron half-light radius as a function of the mass-weighted gas-phase metallicity when adopting the metallicity dependent DTM and a fixed DTM of 0.3. Focusing first on 850 \micron, we find that this ratio is roughly unity for galaxies with a gas-phase metallicity of 0.75 $Z_\odot$ and higher. At lower metallicities the ratio is lower, approximately 0.9 at 0.5 $Z_\odot$ and down to 0.8 at even lower metallicites. At 1.6 \micron, we find that the ratio is approximately 1.1, independent of the gas-phase metallicity of the galaxies. Only at $z=5$ does the model predict ratios up to 1.3.

In summary, when adopting a fixed DTM of 0.3 the model predicts more extended 850 \micron emission and more compact 1.6 \micron emission, all typically within 10-20 per cent from our fiducial model. These results suggest that our conclusions are qualitatively robust against expected changes in the DTM. In fact, we have checked that the claims and quantitative statements about the 850 to 1.6 \micron size ratios of Figure~\ref{fig:StellarRatio}, top right panel, remain essentially unchanged to better than the 10 per cent level.

\end{document}